\documentclass{aa}
\usepackage{graphicx}
\usepackage{natbib}

\usepackage{datetime}
\usepackage[varg]{txfonts}
\newcommand{\HI}{{\ion{H}{I}}}

\newcommand{\kms}{$\,$km$\,$s$^{-1}$}

\newcommand{\mJybeam}{mJy beam$^{-1}$}
\newcommand{\muJybeam}{$\mu$Jy beam$^{-1}$}
\newcommand{\msun}{{$M_\odot$}}

\newcommand{\texci}{$T_{\rm ex}$}

\newcommand{\ltsima} {$\; \buildrel < \over \sim \;$}
\newcommand{\gtsima} {$\; \buildrel > \over \sim \;$}
\newcommand{\lta} {\lower.5ex\hbox{\ltsima}}
\newcommand{\gta} {\lower.5ex\hbox{\gtsima}}

\newcommand{\hco}{HCO$^+$(4-3)}
\newcommand{\dtco}{$^{13}$CO(2-1)}
\newcommand{\twco}{$^{12}$CO(2-1)}
\newcommand{\rde}{$R_{31}$}
\newcommand{\rdt}{$R_{32}$}
\newcommand{\rvt}{$R_{42}$}
\newcommand{\rte}{$R_{21}$}
\newcommand{\nhtwee}{$n_{\mathrm{H_2}}$}
\newcommand{\tkin}{$T_{\mathrm{kin}}$}

\usepackage{color}

\begin{document}

\title{Properties of the molecular gas in the fast outflow\\ in the Seyfert galaxy IC~5063}
\titlerunning{Properties of the molecular gas in the fast outflow in the Seyfert galaxy IC~5063}
\authorrunning{Oosterloo et al.}
\author{
Tom Oosterloo\inst{1,2}, 
J. B. Raymond  Oonk\inst{1,3}, 
Raffaella Morganti\inst{1,2},  
Fran\c{c}oise Combes\inst{4,5}, 
Kalliopi Dasyra\inst{6,7}, \\
Philippe Salom\'e\inst{4}, 
Nektarios Vlahakis\inst{6},
and
Clive Tadhunter\inst{8}}
\institute{
ASTRON, the Netherlands Institute for Radio Astronomy, Postbus 2, 7990 AA Dwingeloo, The Netherlands
\and
Kapteyn Astronomical Institute, University of Groningen, Postbus 800,
9700 AV Groningen, The Netherlands
\and
Leiden Observatory, Leiden University, Postbus 9513, 2300 RA Leiden, The Netherlands
\and
LERMA, Observatoire de Paris, CNRS, UPMC, PSL Univ., Sorbonne Univ., 75014 Paris, France
\and
Coll\`ege de France, 11 place Marcelin Berthelot, 75005, Paris, France
\and
Department of Astrophysics, Astronomy \& Mechanics, Faculty of Physics, National and Kapodistrian University of Athens, Panepistimiopolis Zografou, 15784, Greece
\and
National Observatory of Athens, Institute for Astronomy, Astrophysics, Space Applications and Remote Sensing, Penteli, 15236, Athens, Greece
\and
Department of Physics and Astronomy, University of Sheffield, Hounsfield Road, Sheffield S3 7RH, UK 
}

\offprints{oosterloo@astron.nl}

%\date{Received ...; accepted ...}
%\date{Draft: \today\ -- \currenttime.}
\date{re-submitted: \today}

\abstract{We present a detailed study of the properties of the molecular gas in the fast AGN-driven outflow in the nearby radio-loud Seyfert galaxy IC 5063. By using ALMA observations of a number of tracers of the molecular gas ($^{12}$CO(1-0), $^{12}$CO(2-1), $^{12}$CO(3-2),  $^{13}$CO(2-1) and \hco), we  map the differences in excitation, density and temperature of the gas as function of position and kinematics. The results show that in the immediate vicinity of the radio jet, a fast outflow, with velocities up to 800 \kms, is occurring of which the gas has high excitation with excitation temperatures in the range 30--55 K, demonstrating the direct impact of the jet on the ISM.  The relative brightness  of the   $^{12}$CO lines, as well as that of \dtco\ vs \twco, show that the outflow is optically thin. We estimate the mass of the molecular outflow to be at least $1.2\times 10^6$ \msun\ and likely to be a factor 2--3 larger than this value.  This is similar to that of the outflow of atomic gas, but much larger than that of the ionised outflow, showing that the outflow in IC 5063 is dominated by cold gas.  The total mass outflow rate we estimate to be  $\sim$12 \msun\ yr$^{-1}$. The  mass of the outflow is much smaller than the total gas mass of the ISM of IC 5063.  Therefore, although the influence of the AGN and its radio jet is very significant in the inner regions of IC 5063, globally speaking the impact will be very modest. \\
We use RADEX non-LTE modelling to explore the physical conditions of the molecular gas in the outflow. Models with  the outflowing gas being quite clumpy give the most consistent results and our preferred solutions  have kinetic temperatures in the range 20--100 K and  densities  between  $10^5$ and $10^6$ cm$^{-3}$. The resulting pressures are $10^6$--$10^{7.5}$ K cm$^{-3}$, about two orders of magnitude higher than in the outer quiescent disk. The highest densities and temperatures are found in the regions with the fastest outflow. \\
The results strongly suggest that the outflow in IC 5063  is driven by the radio plasma jet  expanding into a clumpy gaseous medium and creating a cocoon of (shocked) gas which is pushed away from the jet axis resulting in  a lateral outflow, very similar to what is predicted by numerical simulations. 
}

\maketitle  

\begin{figure*}
\centering
\includegraphics[width=\hsize,angle=0]{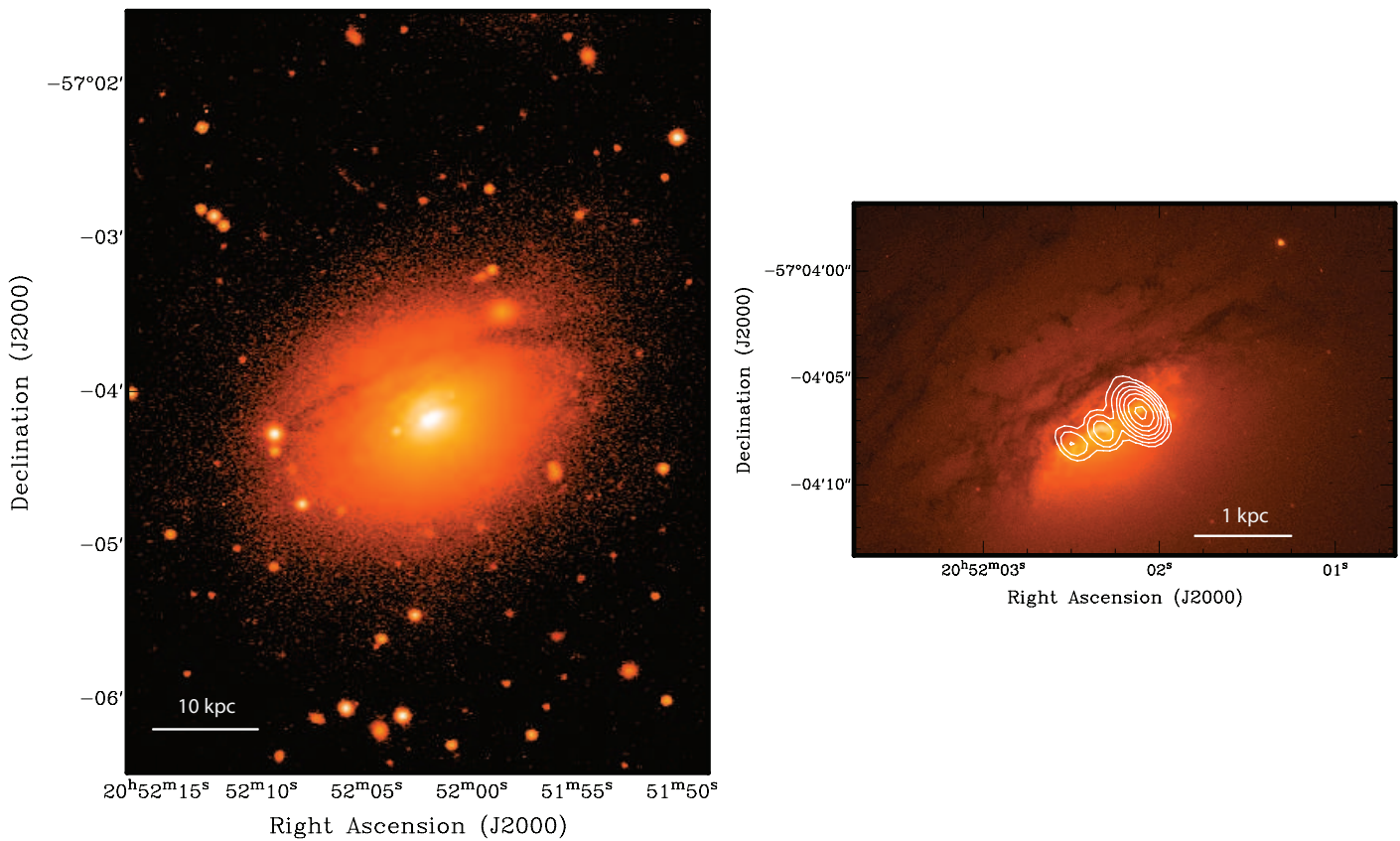}
\caption{ Left: optical image of IC 5063. This image is taken, using EFOSC on the ESO 3.6-m telescope, through a narrow-band filter centred on 5100 \AA,  and shows the large-scale dust lanes and the overall structure of the galaxy \citep{Tsvetanov1997}. Right:   HST image of IC 5063 of the central regions of IC 5063, taken  from the public HST archive. This image is a single 500 s exposure obtained with WFPC2 through the F606W filter, and it shows the  structure of the inner dust lanes in the central region. Overplotted are the contours of the 8 GHz continuum emission of the central kpc-sized radio source   showing the central core and the eastern and western  radio lobes \citep[data from][]{Morganti1998}. Contour levels are 2, 4, 8, 16, 32, 64 and 128 \mJybeam. 
}
\label{fig:optImages}
\end{figure*}

\section{Introduction}
\label{sec:introduction}

The study of fast, galactic scale gas outflows (with velocities exceeding 1000 \kms\ in some cases)  driven by active galactic nuclei (AGN) has become very relevant for understanding the role of  AGN in galaxy evolution because of the  large amounts of energy such outflows can dump into the ISM of the host galaxy or in the gaseous medium surrounding it, thereby  affecting the progress of star formation and of the growth of the central super-massive black hole. Interestingly, such studies have brought a number of surprises about the physical properties of the gas impacted by the AGN.
One of the unexpected results of recent work has been the finding that these fast AGN-driven outflows  contain a large, in terms of mass often dominant,  component of cold (atomic and molecular) gas \citep[see][for a recent compilation]{Fiore2017}.  The presence of large amounts of cold gas in fast outflows is  interesting for at least two reasons: the origin -- how can cold gas survive, or form, in fast outflows -- and whether fast outflows not only quench star formation, but, instead, also can, under certain circumstances, induce star formation \citep[e.g.][]{Silk2013,Maiolino2017}.

Despite the  relevance of fast outflows, detailed studies of cold, fast outflows are still limited to a few objects, and, given the diversity of the characteristics of these objects, our understanding of the typical physical conditions of the cold gas is limited. Improving our knowledge of the properties of cold gas in such harsh environments is key for developing physical models for the outflows and what drives them, as well as for assessing the overall impact of the AGN, and the outflows it drives, on galaxy evolution.

The limited number of objects where molecular outflows not only have been detected, but in addition also have their physical properties investigated using multiple line transitions, show  large differences in physical conditions between the quiescent, non-affected  gas and the outflowing gas, underlining the impact of the driving mechanism   (e.g.\ NGC 1068, \citealt{Viti2014,Garcia2014}; Mrk 231, \citealt{Aalto2012,Cicone2012,Feruglio2015}; NGC 1266, \citealt{Alatalo2011}; 4C12.50, \citealt{Dasyra2014} and NGC 1433, \citealt{Combes2013}). The results indicate that the outflowing molecular gas is clumpy and has much higher temperatures and densities than the molecular gas in the normal ISM. The picture that emerges is that the gas is compressed and fragmented by the mechanism that drives the outflow.
% The results also show that the outflowing gas is optically thin, which has implications for deriving properties of the outflow, in particular its mass.

One limiting factor is that not many objects are known for which one can do a detailed, {\sl spatially well resolved} analysis of the properties and kinematics of the outflow. One of the few objects for which this is possible is the nearby Seyfert galaxy IC 5063. This  was the first object where a fast AGN-driven outflow of {\sl atomic} hydrogen was discovered \citep{Morganti1998}. IC 5063 is an early-type galaxy ($z = 0.1134$\footnote{Assuming a Hubble constant
  $H_{\circ}$= 70 km s$^{-1}$ Mpc$^{-1}$, $\Omega_\Lambda=0.7$ and
  $\Omega_{\rm M} = 0.3$, the redshift of IC 5063 ($V_{\rm hel}$ = 3400 \kms) implies an angular size distance of 47.9 Mpc and a scale of 1 arcsec = 232 pc.})  with a number of prominent dust lanes (see Fig.\ \ref{fig:optImages}). The galaxy is gas rich, having a large-scale, regularly rotating gas disk containing $4.2 \times 10^9$ \msun\ of atomic hydrogen \citep{Morganti1998} , and at least $10^9$ \msun\ of molecular hydrogen \citep{Morganti2015}.

IC 5063 is one of the most radio-loud Seyfert galaxies known (albeit,  in a general sense, still  a relatively weak radio AGN, with power $P_{\rm 1.4 GHz} = 3\times10^{23}$ W Hz$^{-1}$). The radio continuum emission comes from  a linear triple structure, aligned with the inner dust lane, consisting of a central core and two lobes  \citep{Morganti1998}, of about 4 arcseconds in size (corresponding to $\sim$1 kpc; see Fig.\ \ref{fig:optImages}). Observations of the atomic hydrogen in IC 5063 quite unexpectedly revealed the presence of a fast outflow, with outflow velocities up to 700 \kms\ \citep{Morganti1998}. 
Subsequent  VLBI observations of the \HI\ in IC 5063 \citep{Oosterloo2000}, as well as observations of the ionised gas \citep{Morganti2007,Dasyra2015}, suggested the likely dominant role of the radio jet, and the over-pressured cocoon surrounding it, in driving the outflow. The presence of molecular gas associated with this outflow was first found from APEX CO(2-1) observations \citep{Morganti2013} and by the detection of warm molecular gas (H$_2$ detected at 2.2 $\mu$m) using ISAAC on the VLT \citep{Tadhunter2014}. The latter study also showed that the warm H$_2$ gas with the most extreme outflow velocities is co-spatial with the bright radio hot-spot $\sim$0.5 kpc west of the nucleus, underlining the importance of the jet in driving the outflow. 

However,  the high spatial resolution  ALMA CO(2-1) observations   presented in  \citet{Morganti2015}  give the best view of the complex interplay between the radio plasma ejected by the AGN and the molecular gas in the central regions of IC 5063. These observations have shown the presence of a central, bright component of CO(2-1) emission ($\sim$1 kpc in size)  which has a close spatial correspondence with the radio jet \citep[see Fig.\ 1 in][]{Morganti2015}.  In addition, this bright   molecular  gas has  strongly disturbed kinematics, extending along the entire radio jet with outflow velocities up to 800 \kms\  with the highest outflow velocities occurring about 0.5 kpc from the nucleus, at the location of the bright hot-spot in the W lobe. The close spatial correspondence of this region of molecular gas having highly disturbed kinematics with the region of radio plasma, suggests  a model -- similar to, e.g.,  presented in the simulations of \citet{Wagner2012} -- where the radio plasma jet is expanding into a clumpy gaseous medium and it creates a cocoon of (shocked) gas which is pushed away from the jet axis resulting in  a lateral expansion.  In the simulations of \citet{Wagner2012}, different zones in the gas affected by the AGN can be identified and, according to the model, different conditions must occur in these different zones. Thus, given that the molecular outflow in IC 5063 is spatially well resolved,  this object is an ideal candidate for studying the physical conditions of the molecular outflows and to compare observations with models.  A detailed comparison of our data with numerical models will be done in a following paper  \citep{Mukherjee2017}.

First results were presented in \citet{Dasyra2016} where new ALMA observations of the CO(4-3)  transition were combined with the CO(2-1) data of \citet{Morganti2015}. This combination  showed that the physical conditions of  the molecular gas in the jet-driven outflow are very different from those of the larger-scale quiescent molecular disk, with the former having much higher excitation temperatures. Importantly, the observed  line ratios indicated that the outflowing gas is optically thin. This has important implications for how to convert the observed line intensity of the outflow to its mass and  results in a lower estimate of the mass of the outflow compared to the optically thick case.

Here, we further expand on these results by presenting new ALMA observations exploring other tracers of the  molecular gas:  $^{12}$CO(1-0), $^{12}$CO(3-2),  $^{13}$CO(2-1) and \hco. This larger range of molecular tracers allows more extensive  characterisation of the physical conditions of the different kinematical components    of the molecular gas. The structure of the paper is as follows: in Sect.\ 2 we describe the ALMA observations and the data reduction. In Sect.\ 3 we compare the spatial distribution and kinematics of the $^{12}$CO lines, while in Sect.\ 4 we present the \hco\ and $^{13}$CO observational results. In Sect.\ 5 we discuss the excitation of the gas in the various regions of the outflow and derive an estimate of the mass of the molecular outflow. In this section we also present estimates of the kinetic temperatures and densities for representative regions in the outflow and in the disk of IC5063 derived using the RADEX non-LTE radiative transfer code\footnote{https://personal.sron.nl/$\sim$vdtak/radex/index.shtml}  \citep{Tak2007}.

\section{Observations and data reduction}

\begin{figure*}
\centering 
\includegraphics[width=0.9\hsize]{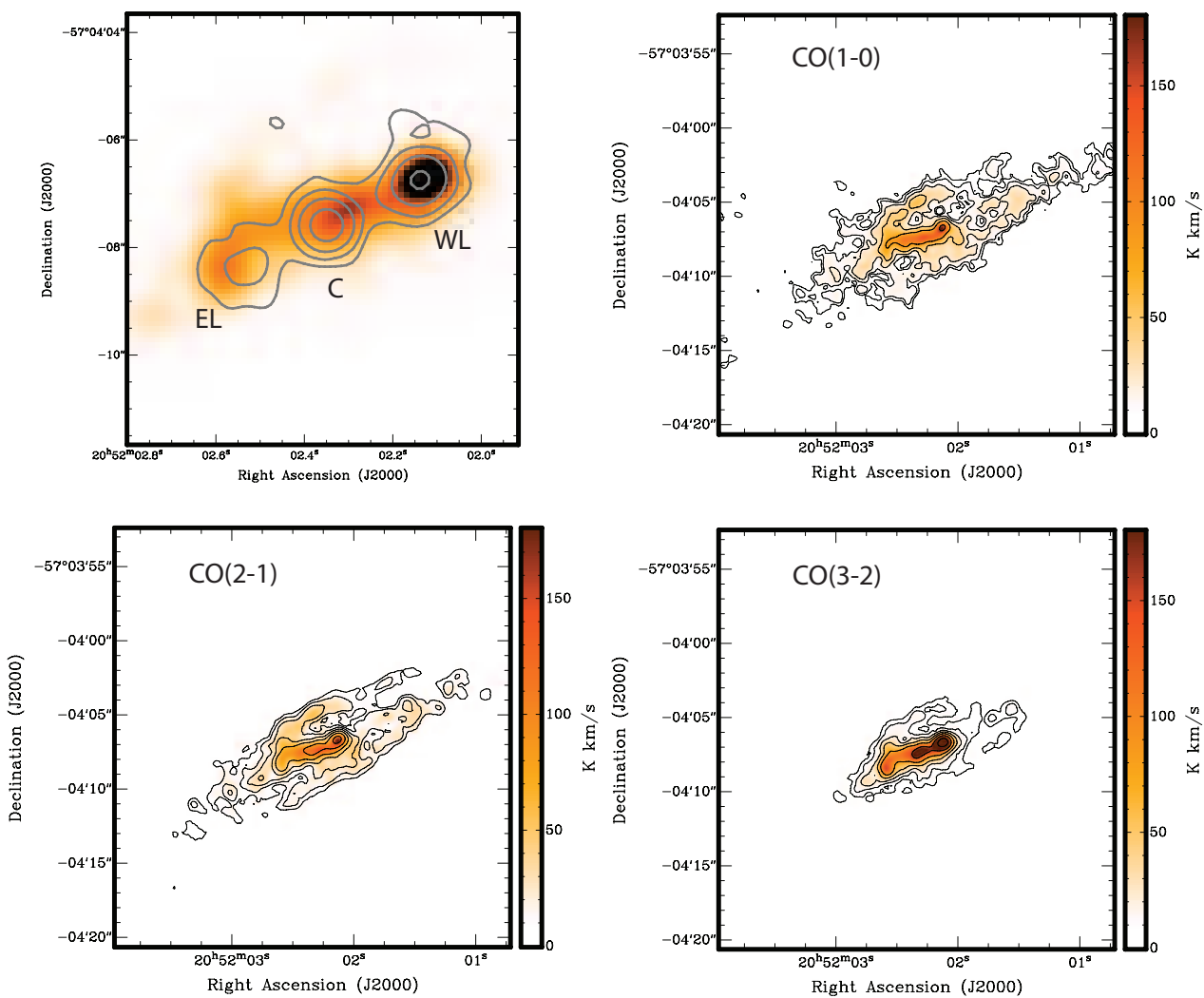}
\caption{ Top left: integrated brightness temperature of the CO(3-2) emission  (grey scale) with superposed the contours of the 346 GHz continuum, to illustrate the close spatial correspondence between the radio continuum and the bright, inner molecular structure.  The core (C) and the eastern and western lobe (EL and WL respectively) are indicated.  Contour levels are  0.15, 0.45, 1.35 and 4.05 \mJybeam. The other panels show the integrated brightness temperature  for CO(1-0) (top right), CO(2-1) (bottom left) and CO(3-2) (bottom right). Contour levels are 5, 10, 20, 40,... K \kms\ for these three panels, while also the colour stretch is the same for these three images.}
 \label{fig:totImages}
\end{figure*}

The ALMA data discussed in this paper come from a number of observing sessions, while we also re-use our earlier ALMA observations of the CO(2-1) emission of IC 5063 published in \citet{Morganti2015}. Since  only the 12-m array was used and observations at very different frequencies are involved, we have taken care to use the proper array configurations for the different observations.  This was done to ensure that the various observations have very similar spatial resolution, while in addition the  $uv$ coverages  are as similar as possible such that sensitivity to different spatial scales is the same at different frequencies.  No signatures of missing large-scale flux ('negative bowl') were present in any of the observations. 

The CO(3-2) and \hco\ data were taken with Band 7 on March 22, 2016, using 37 antennas. The frequency setup of the telescope employed four observing bands. Two high-resolution spectral windows, each having a total bandwidth of 1.875 GHz split over 480 channels,  were used to image the CO(3-2)  (centred on 341.97 GHz) and  the \hco\ emission (centred on 353.05 GHz) of IC 5063. In addition, two continuum spectral windows were used, each 2 GHz wide and having 128 channels, centred on 340.1 and 352.0 GHz. The total on-source observing time was about 0.5 hr. Standard calibration observations were done, involving Titan, J2056$-$4714 and J2141$-$6411 as flux, bandpass and phase calibrators respectively. 

To study the properties of the \dtco\ emission in IC 5063, we used ALMA  in Band 6 on May 19 and 23, 2016, employing 38 and 35 antennas on the respective days. The total on-source observing time is 2.5 hr.  The calibration strategy is the same as  for the Band 7 observations, except that Pallas was used as a flux calibrator. Two 2-GHz wide continuum spectral windows were used (at 230.0 and 232.5 GHz, each band covered with 128 channels) as well as a higher resolution spectral window centred on 217.8 GHz (1.875 GHz wide, using 960 channels).

Finally, the CO(1-0) emission of IC 5063 was observed with ALMA in Band 3 on July 25 and 26, 2016 using 45 antennas. The total on-source time was 1.9 hr. High spectral resolution data were taken using a 1.875 GHz wide band (960 channels) centred on 114.0 GHz. Continuum spectral windows of 2 GHz wide centred on  100.1 and 101.9 GHz were used to image the continuum emission of IC 5063 in this band. Also for these observations, standard calibration was done, involving the same sources as for the \dtco\ observations.

For all observations, the initial calibration of the data was done in CASA\footnote{http://casa.nrao.edu} \citep{McMullin2007}) using the calibration scripts and calibration data provided by ESO. We found, however, that we were able to significantly improve the  dynamic range of the images and cubes by additional self calibration (phase only) of the  continuum images of IC 5063 and apply these calibrations also to the line data. This self calibration and all subsequent imaging and data reduction were done using the MIRIAD\footnote{http://www.atnf.csiro.au/computing/software/miriad} software \citep{Sault1995} as well as self-made PYTHON scripts.

Our main aim is to study the relative strength of the different line transitions in different regions of IC 5063. Hence, when making the images and data cubes, care has to be taken that they have the same spatial and velocity resolution. In addition, it is well known that self calibration can introduce small position offsets in images. Therefore, we used the continuum images to determine possible offsets between data sets after the self calibration. We indeed found that the self calibration had introduced offsets of the order of 0.1 arcsec.  All images and data cubes we present are corrected for these offsets.

The data cubes of the main CO transitions were made using the same spatial (0\farcs1 pixels) and velocity gridding (10 \kms\ channels, but Hanning smoothing was applied to the data cubes so that the velocity resolution is 20 \kms). From all three observations, data cubes were made using natural weighting. Although the $uv$ coverages of the different observations are very similar, they are not identical. This resulted in small differences (of about 0\farcs05) in the resolution of the data cubes of the different transitions. We corrected the data cubes for this by smoothing the data cubes to a common resolution. This resulted in data cubes having a resolution of 0\farcs62. This corresponds, for the distance to IC  5063 assumed, to a linear resolution of 143 pc.  %(PA 64\fdg7). 
  
All cubes were cleaned using an iterative masking technique using a lower resolution version of a cube of a previous iteration to mask emission regions to be cleaned in the high-resolution cube in the next iteration. This ensures that the emission can be cleaned below the noise level while noise peaks are not included in the cleaning process. The noise level of the CO(1-0) data cube is 0.30 \mJybeam\ (corresponding to 73.0 mK), of the CO(2-1) data cube  0.23 \mJybeam\ (13.6 mK)  and that of the CO(3-2) data cube  0.55 \mJybeam (15.0 mK).

The \hco\ and the \dtco\ data were initially treated in the same way as those of the main CO transitions and data cubes were made with the same spatial and velocity resolution as described above. However, as expected, the \hco\ and \dtco\ emission is much fainter than that of the main CO lines and in the full-resolution cubes the emission is detected only close to the noise level. Therefore, in order to improve the signal-to-noise of the emission, we tapered the data to much lower spatial and velocity resolution in order to better match the spatial and velocity structure of the emission. The \hco\ and \dtco\ cubes we  discuss have a spatial resolution of 1\farcs52 (corresponding to about 350 pc) and a velocity resolution of 100 \kms. The noise level of the \hco\ cube is 0.90 \mJybeam\ (3.8 mK) and of the \dtco\ cube 0.18 \mJybeam\ (2.0 mK).  To enable comparison, data cubes with the same low spatial and velocity resolution were also made for the main CO transitions. 

To study the relation between the molecular gas and the radio continuum emission, a continuum image was made  (Fig.\ \ref{fig:totImages}) by combining the continuum spectral windows observed at 340.1 and 352.0 GHz using multi-frequency synthesis (giving an effective frequency of about 346 GHz).  Also here, natural weighting was used for gridding the $uv$ data. The noise level of this continuum image is 50 \muJybeam\ and the spatial resolution is  0\farcs60.  This continuum image shows the same triple structure seen at  8 GHz \citep[][see also Fig.\ \ref{fig:optImages}]{Morganti1998}, 17.8 and 24.8 GHz \citep{Morganti2007} and 230 GHz \citep{Morganti2015}. The central peak is the core of radio source which connects through  jets to the eastern and western lobes. The great similarity between the continuum images over this wide range of frequencies indicates that the bulk of the radio continuum at  346 GHz is non-thermal emission from the AGN and the radio lobes.

In the following,  we first compare the distribution and kinematics of the various lines observed to study  the overall impact of the radio jet on the molecular gas in the centre of IC 5063. This is followed by a more detailed analysis of the physical conditions of the gas in various regions in IC 5063.

\section{Comparison between the main CO lines}

\subsection{Spatial distribution}

One of the main aims of this paper is to further study  the properties of the  molecular gas participating in the fast  outflow in IC 5063. In \citet{Dasyra2016} we found marked differences in the line ratio  \rvt\ $\equiv I_{4-3}/I_{2-1}$ between the outer disk and the inner regions. At many locations in  the inner regions, \rvt\ takes values well above 1, indicating optically thin conditions as well as excitation temperatures in the range 30--50 K and in some locations even higher. On the other hand, the low values found for \rvt\ in the outer disk indicates sub-thermal excitation of the molecular gas there.  The large contrast between the two regions signifies the impact of the radio jet on the gas in the central regions of IC 5063. Here, we will do a similar analysis of the conditions in the inner regions using our new observations. 

As a first approach, we compare the sky distributions  of the  integrated intensity of the main CO   transitions which are shown in Fig.\   \ref{fig:totImages}. These images were made in a standard way, i.e.\ by masking the full-resolution data cubes  using a 2-$\sigma$ clipped lower resolution version of the cubes, followed by integrating the masked, high-resolution cube over velocity.

\begin{figure}
\centering
\includegraphics[width=\hsize,angle=0]{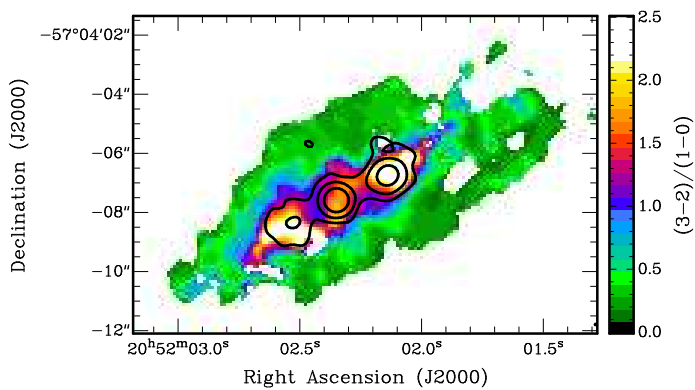}
\caption{Ratio $R_{31}$ of the CO(3-2) and CO(1-0) integrated brightness temperatures $I$ (in K \kms) of Fig.\ \ref{fig:totImages}, after primary beam correction. Only those pixels are shown for which the error in \rde\ is smaller than 0.3. Overplotted are the contours of the 346 GHz continuum emission of IC 5063 with contour levels 0.15 (3$\sigma$), 0.6,  2.4 and 9.6 \mJybeam. 
}
\label{fig:ratImage}
\end{figure}

Our earlier data \citep{Morganti2015,Dasyra2016} had already shown that the distribution of the molecular gas in IC 5063, as observed in the CO(2-1) and CO(4-3) lines,  consists of a bright inner structure tightly surrounding the bright radio continuum emission of the core and radio lobes,  and a more extended, lower surface brightness  molecular counterpart of the large \HI\ disk of IC 5063. Figure \ref{fig:totImages} shows that, as expected, this is also the case for the other CO transitions. Figure \ref{fig:totImages} also confirms the large  difference of the relative brightness of the bright inner and the fainter outer regions  in the different transitions with the inner region  being relatively much brighter for the higher transitions. This is in line with the even larger contrast between the two components seen in the comparison of the CO(2-1) and CO(4-3) data presented in \citet{Dasyra2016}.  

To  quantify this difference in contrast, we have computed, using the images given in Fig.\ \ref{fig:totImages}, the spatial distribution of the ratio \rde $\equiv I_{3-2}/I_{1-0}$ (with $I$ in K \kms; see Fig.\ \ref{fig:ratImage}). 
The image of \rde\ clearly confirms that there are two distinct regions in  IC 5063: an inner region  with values of \rde\ well above 1, and an outer region with  values below 0.5. A similar pattern is seen in images (not shown here) for the ratio \rdt\ $\equiv I_{3-2}/I_{2-1}$ where \rdt\ is around 1.0 or somewhat larger in the inner region and has values in the range 0.3--0.6 in the outer disk.

Figure \ref{fig:ratImage}  clearly shows that, either directly or indirectly, the radio jet in IC 5063 must be responsible  for the very different conditions of the gas in the inner regions. In \citet{Morganti2015} and \citet{Dasyra2016} we had already observed that the bright, inner CO structure is spatially closely connected to the   radio continuum and interpreted this as  that the inner CO is bright because the radio jet strongly affects the excitation of the gas in the inner regions. Our new image of \rde\  shows this in dramatic form: it is exactly the bright inner region where the high values for the line ratios are found and the region with high values of \rde\  very nicely encompasses the radio continuum.   This is very similar to what seen in, e.g., NGC 1068  where the  region with high CO line ratios is also closely connected to the region of the AGN \citep{Garcia2014}. 

In addition, Fig.\ \ref{fig:ratImage}  shows that the locations with the highest line ratios (with values well above 2.0)  exactly coincide with the two lobes. We also note  that the transition between the region with the high values of \rde\  to the region with lower values at larger radii is very sharp. This underlines that the  region with very different conditions compared to the outer, quiescent disk, is  limited to the  direct environment of the radio jet.

We will not further discuss the molecular gas in the outer quiescent disk, other than that  we note that \rde\   in these regions  has quite low values, in the range 0.3--0.6. We find similar values in our data for, e.g., \rdt\ $ \equiv I_{3-2}/I_{2-1}$.  This is consistent with \rvt\ being $\sim$0.25 as we found in \citet{Dasyra2016} for the same region. The molecular gas in the star forming disks of spiral galaxies typically show values for  $R_{21} \equiv I_{2-1}/I_{1-0}\sim 0.8$ \citep[e.g.,][]{Leroy2009} implying higher values for \rde\ and \rdt\ than seen here in the outer disk of IC 5063. Beyond the central regions, the gas density of the ISM in the disk of IC 5063, as well as the amount of star formation,  is lower than that typically seen in star forming spiral galaxies \citep{Morganti1998} and the low values of \rde,  \rdt\  and \rvt\ in  the disk of IC 5063 are likely due to, as suggested in \citet{Dasyra2016},  the excitation of the molecular gas in the outer regions in IC 5063 being sub-thermal.

\begin{figure}
\centering
\includegraphics[width=\hsize,angle=0]{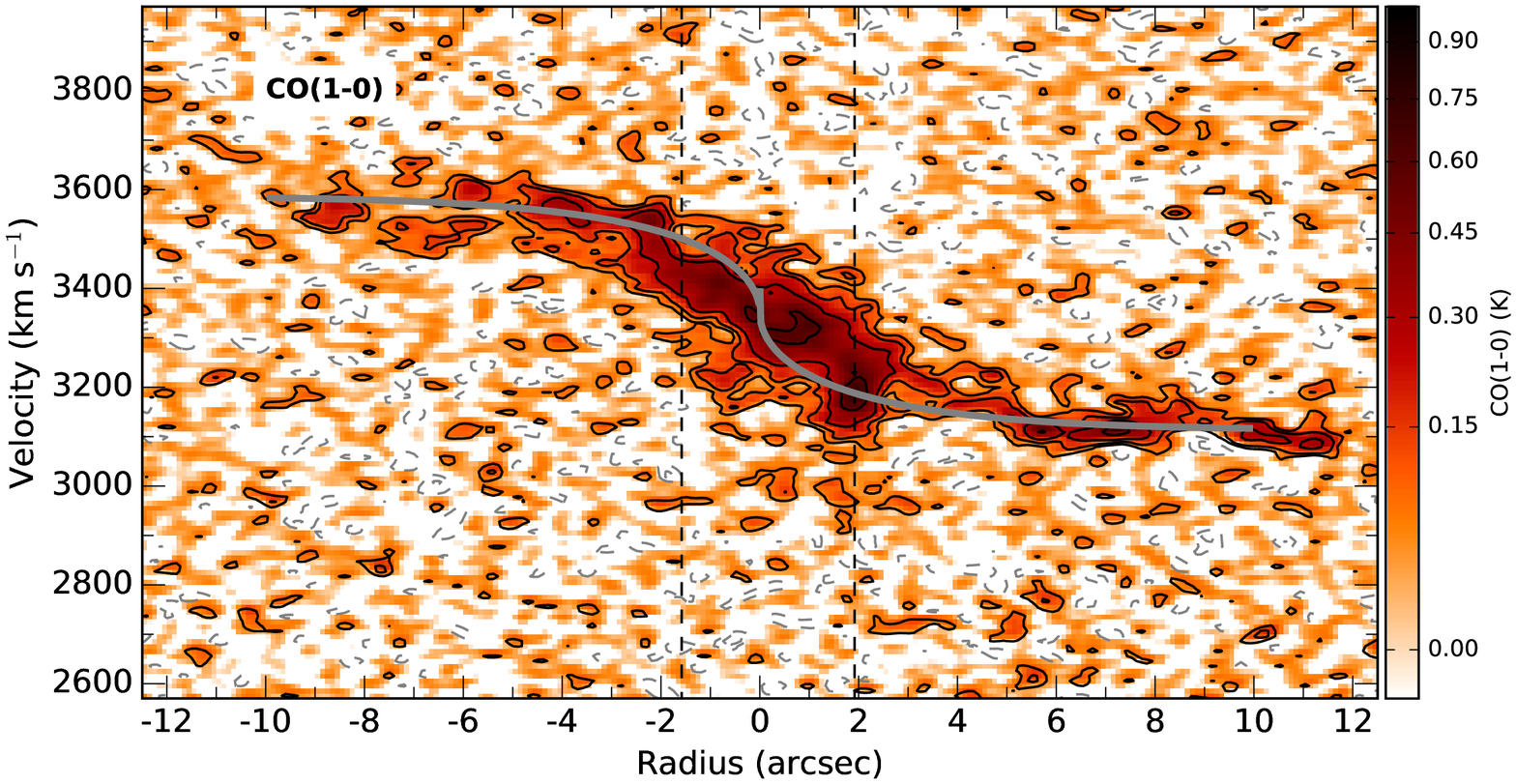}
\includegraphics[width=\hsize,angle=0]{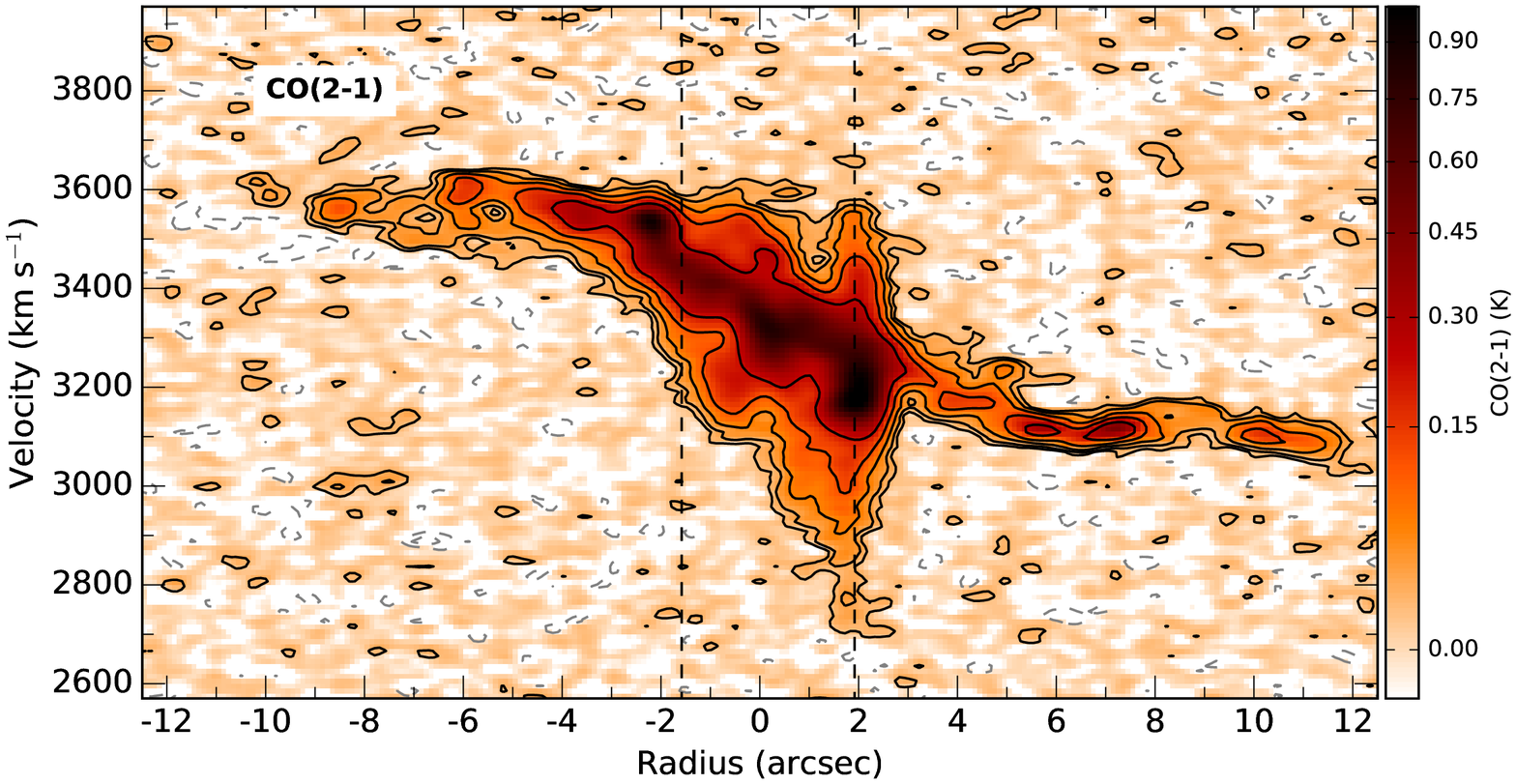}
\includegraphics[width=\hsize,angle=0]{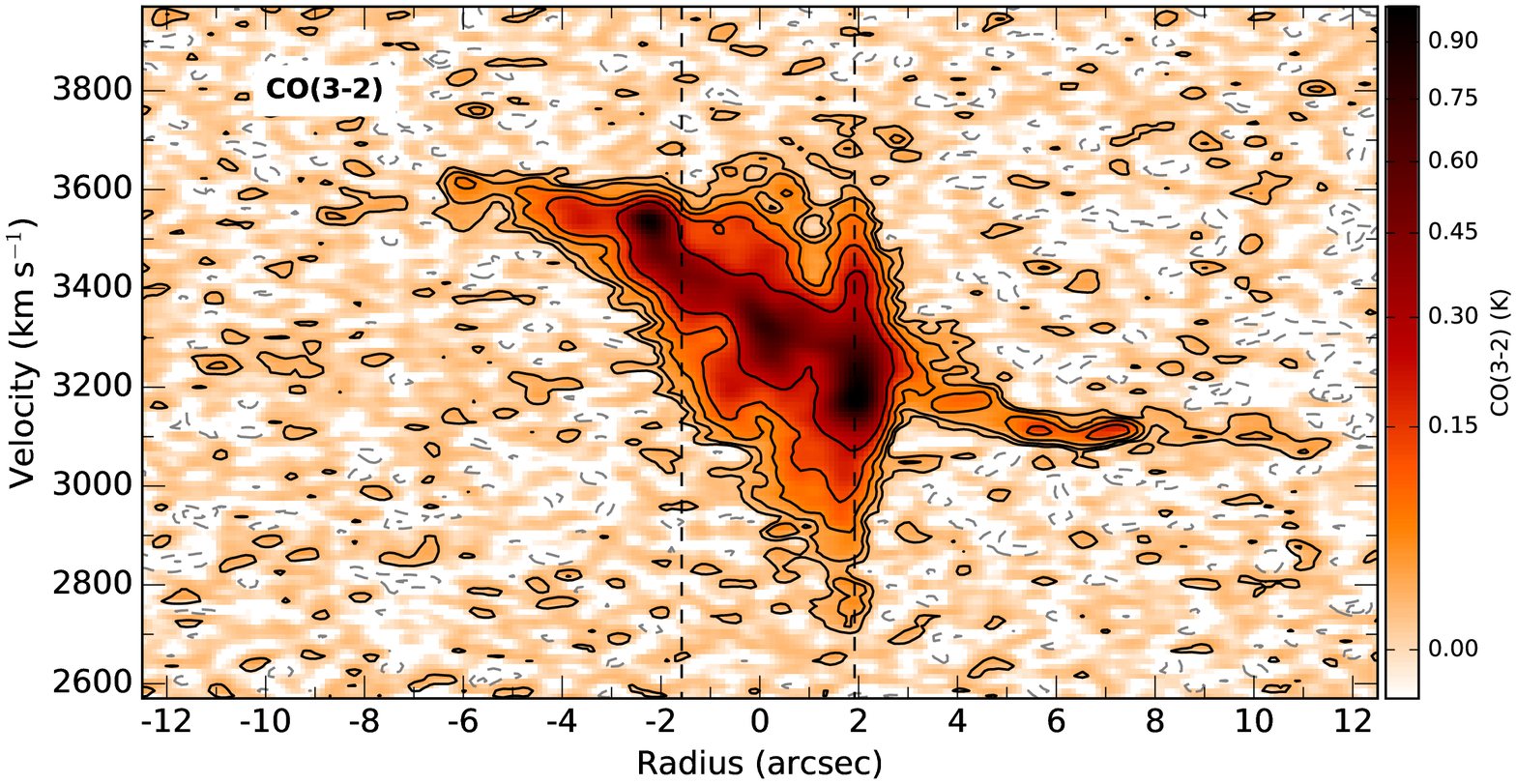}
\caption{Position-velocity plots along the jet axis for CO(1-0) (top), CO(2-1) (middle) and CO(3-2) (bottom) where the data  have been averaged over 1.5 arcsec perpendicular to the jet axis. Contour levels are --71.25 (dashed),  71.25   (1.5 $\sigma$), 142.5,     285.0, \ldots\ for the CO(1-0) data and --15 (dashed), 15 (1.5 $\sigma$), 30, 60, \ldots\  mK for the other two panels. For all panels the colour stretch is the same.  For display purposes, the data have not been corrected for the primary beam. The dashed vertical lines indicate the approximate extent of the radio continuum structure. Positive radii are west of the centre of IC 5063. The spatial resolution along the horizontal axis is 0\farcs63.  The grey curve  in the top panel indicates  where one would  expect in this $pv$ diagram to detect gas which follows the normal kinematics of a regularly rotating gas disk (i.e.\ gas not participating in the outflow; see text).} 
\label{fig:intSlices}
\end{figure}

\begin{figure}
\centering
\includegraphics[width=\hsize,angle=0]{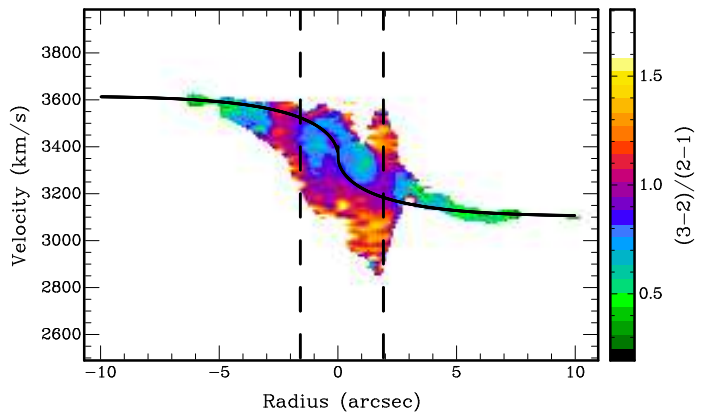}
\caption{Ratio \rdt\  of the CO(3-2) and CO(2-1) brightness temperatures along the jet axis of IC 5063 where the data, before taking the ratio, have been averaged over 1.5 arcsec perpendicular to the jet axis (i.e.\ using the data presented in Fig.\ \ref{fig:intSlices} after applying the primary beam corrections). Only ratios with error smaller than 0.3 are shown.  The black curve indicates  where one would  expect in this $pv$ diagram to detect gas which follows the normal kinematics of a regularly rotating gas disk (i.e.\ gas not participating in the outflow; see text). The black dashed vertical  lines indicate the extent of the radio continuum. }
\label{fig:ratSlice}
\end{figure}

\subsection{Kinematics of the main CO lines}

The results of the previous section clearly show the spatial relationship between the jet and the conditions of the molecular gas. However, because $R_{31}$ as discussed above is derived after integrating the CO emission over velocity, no kinematical information  can be obtained. To study how the gas conditions depend on the kinematics of the gas,   we produced position-velocity ($pv$) diagrams of the CO(1-0), CO(2-1) and (3-2) emission by  averaging  the CO data cubes  over a width of 1\farcs5 (corresponding to 350 pc) in the direction perpendicular to the jet axis (which we have taken to have position angle  $-65^\circ$). The resulting $pv$ diagrams are shown in Fig.\ \ref{fig:intSlices}.  The kinematics of the CO gas, in particular the fast outflow in the inner regions, and the connection to the radio jet in IC 5063, we have discussed extensively in \citet{Morganti2015} and \citet{Dasyra2016}  and we only briefly comment on it here. 

 To give a rough indication where one would  expect in this $pv$ diagram to detect gas which follows the normal kinematics of a regularly rotating gas disk (i.e.\ gas not participating in the outflow), we have, as we did in \citet{Morganti2015}, indicated  a rotation curve based on the HST photometry of IC 5063 of \citet{Kulkarni1998}. This photometry shows that the light distribution can be accurately described using a de Vaucouleurs' law. The rotation curve plotted is based on the assumption that light traces mass and that the mass distribution is spherical. These assumptions are most likely not entirely correct and therefore the rotation curve plotted is only approximate, but it suffices to illustrate the anomalous kinematics of the gas in the outflow. The horizontal scale of the rotation curve is set by the observed effective radius of the light distribution (21\farcs5 corresponding to 5.0 kpc), while  the amplitude of the rotation curve is chosen such  to match the rotation velocities of the outer disk. Since IC 5063 is observed almost edge on, regularly rotating gas is expected to lie in the regions with velocities between the plotted rotation curve and the systemic velocity ($V_{\rm hel} = 3400$  \kms, \citealt{Morganti1998}). Gas found in other regions of the $pv$ diagram has "anomalous kinematics", indicating non-circular motions which in this case mean that the gas is participating in the gas outflow.

The $pv$ diagrams of the CO(2-1) and the CO(3-2) emission in Fig. \ref{fig:intSlices} clearly show the fast molecular outflow, i.e.\ the presence of gas, for radii smaller than 2 arcseconds and exactly coinciding with the extent of the radio source, of which the kinematics strongly deviates  from that expected for a regularly rotating disk, with differences up to 700 \kms. In many locations, the velocities of the gas even have the wrong sign compared to those of a regular disk, in particular near the western lobe where the fastest outflow velocities are seen.  We also note the dramatic difference and sharp transition between the kinematics of the  molecular gas  in the region coinciding with the radio continuum ($r <2^{\prime\prime}$) and the regular rotation of the large-scale quiescent  disk beyond this radius. 

The diagrams presented in Fig.\ \ref{fig:intSlices} show that, similarly to those of Fig.\ \ref{fig:totImages}, the inner, bright CO structure, including the fast outflow, is detected in an almost identical way in CO(2-1) and CO(3-2).  The bright, inner ridge is also detected in CO(1-0), but the fast outflowing gas is much fainter (even considering the difference in sensitivity of the different observations).  Below, we  study  in  detail the flux ratios between the various lines we have detected in various locations, but to illustrate the main result,  in Fig.\ \ref{fig:ratSlice} we show \rdt\  as computed from the $pv$ diagrams described above.

The striking  feature in Fig.\ \ref{fig:ratSlice} is that   there is  a strong correlation between \rdt\  and whether the velocities of the gas are anomalous or not. In the inner region, at velocities consistent with those of a regularly rotating disk and not coinciding with the lobes, \rdt\  has values in the range 0.7--0.8, while for gas with velocities well outside the expected range for a regular disk (i.e.\ the gas participating in the fast outflow), as well as the gas coinciding with the lobes,  \rdt\ is in the range 1.0--1.5. The regions with the highest outflow velocities, e.g., for radii between 0 and 2 arcsec and velocities below the rotation curve or above the systemic velocity, \rdt\ is highest.

In Fig.\ \ref{fig:ratImage} where we show \rde\ for the data  integrated over velocity, it appears that the highest line ratios are found near the two radio lobes. However, Fig.\ \ref{fig:ratSlice} makes clear that this is, at least partially, due to mixing different components having different  line ratios, leading to averaging of regular and jet-disturbed gas over velocity. Figure \ref{fig:ratSlice} shows that all along the radio jet,  the emission  with significant anomalous velocities  has similar values  for the line ratio as the gas near the lobes. This suggests that the highest excitation not only occurs in the direct vicinity of the lobes, but also between the core and the lobes, and not only for gas with large outflow velocities. The presence of gas with high excitation but relatively low {\sl apparent} velocities   indicates that the outflow is also occurring in the plane of the sky, i.e.\  perpendicular to the large-scale disk of IC 5063, as was also observed by \citet{Dasyra2015}.    The kinematics of the inner region with small line ratios is consistent with being the inner continuation of the outer disk and we may well see a mix of outflowing gas with large line ratios and regular disk gas with small ratios.  The kinematical superposition of two such components was already recognised in \citet{Dasyra2016}.

\section{HCO$^+$ and $^{13}$CO(2-1)}

In an attempt to obtain more information about the physical conditions of the molecular gas in the inner regions of IC 5063 (see Sec.\ \ref{sec:discussion} for details), we also observed the galaxy in the J=4-3 transition of HCO$^+$ as well as in the  $^{13}$CO(2-1) line. As mentioned above, emission is detected in these two lines, but of course at a much lower level compared to the main CO transitions and we smoothed the data to lower spatial and velocity resolution in order to better match them to the spatial and velocity structure of the emission. The HCO$^+$  and $^{13}$CO(2-1) cubes we discuss here have a spatial resolution of 1\farcs52 and a velocity resolution of 100 \kms.

\begin{figure}
\centering
\includegraphics[width=\hsize,angle=0]{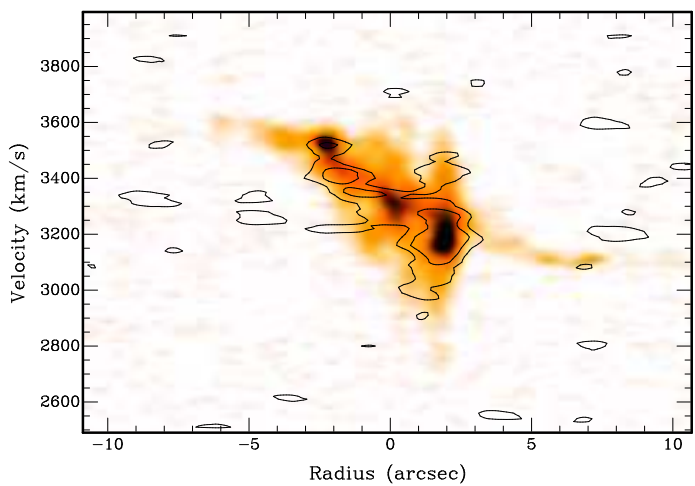}
\caption{Position-velocity diagram of the low-resolution  \hco\ emission (contours)
 along the jet axis of IC 5063 where the data have been averaged over 1.5 arcsec perpendicular to the jet axis.    The greyscale is the position-velocity slice of the full-resolution CO(3-2) data. Contour levels are  6.6 (2$\sigma$) and 13.2 mK.  The spatial resolution along the jet axis is 1\farcs52 and the velocity resolution of the \hco\ data is 100 \kms.}
\label{fig:sliceHCO}
\end{figure}

\begin{figure}
\centering
\includegraphics[width=\hsize,angle=0]{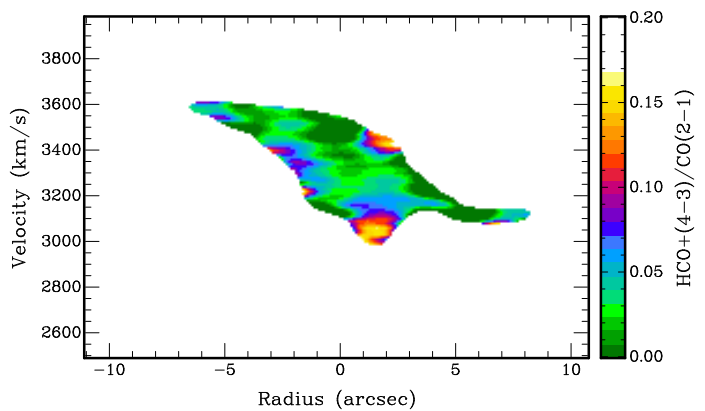}
\caption{Position-velocity diagram of the ratio of the low-resolution \hco\ and CO(2-1) brightness temperatures along the jet axis of IC 5063 where the data, before taking the ratio, have been averaged over 1.5 arcsec perpendicular to the jet axis.  Only pixels for which the error in the ratio is smaller than 0.07 are shown.  The spatial resolution along the jet axis is 1\farcs52. }
\label{fig:ratHCO}
\end{figure}

Figure \ref{fig:sliceHCO} shows the position-velocity diagram of the \hco\ emission obtained in the same way as described above  for the main CO lines (i.e.\ the spatial axis is along the radio jet). The superposition on the CO(3-2) emission shows that the bulk of the \hco\ emission   comes from the bright, inner ridge detected in the main CO lines (which is likely a mix of disturbed and regular gas), but that some faint emission is detected from the fast outflow near the W lobe. Figure \ref{fig:ratHCO} shows the spatial distribution of the ratio of the low-resolution \hco\ and CO(2-1) brightness temperatures in the position-velocity diagrams where only those pixels are displayed that have an error in the line ratio smaller than 0.07. Although  the \hco\ emission is detected at very faint levels and the errors in the ratios are relatively large, Fig.\ \ref{fig:ratHCO} nevertheless suggests that the line ratio of the gas of the outflow is substantially higher than of the more quiescent gas. Given the high critical density of the \hco\ line ($1.8\times 10^6$ cm$^{-3}$; \citealt{Greve2009}) this indicates that the gas  is much denser  in the fast outflow. Relatively strong emission from HCO$^+$ (and other high-density indicators such as HCN) has also been observed in other outflows, such as NGC 1068 \citep{Garcia2014} and Mrk231 \citep{Aalto2012,Cicone2012,Feruglio2015}, and IC 5063 conforms to this rule.

\begin{figure}
\centering
\includegraphics[width=\hsize,angle=0]{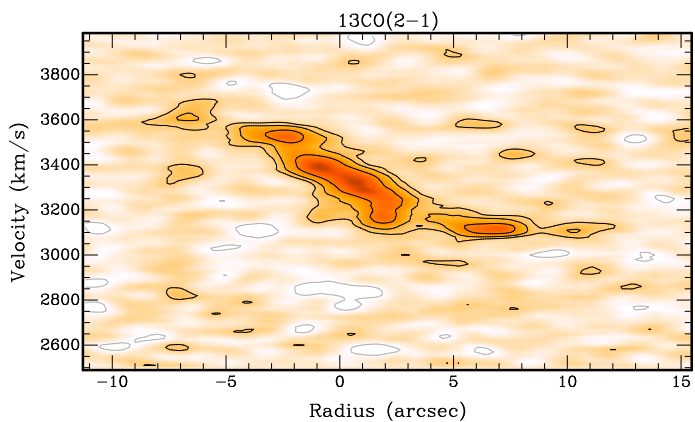}
\caption{Position-velocity diagram of the low-resolution  \dtco\ emission 
 along the jet axis of IC 5063 where the data have been averaged over 1.5 arcsec perpendicular to the jet axis.    Contour levels are --3.6, 3.6 (2$\sigma$), 7.2 and 14.4 mK. The spatial resolution along the jet axis is 1\farcs52. }
\label{fig:slice13CO}
\end{figure}

\begin{figure}
\centering
\includegraphics[width=\hsize,angle=0]{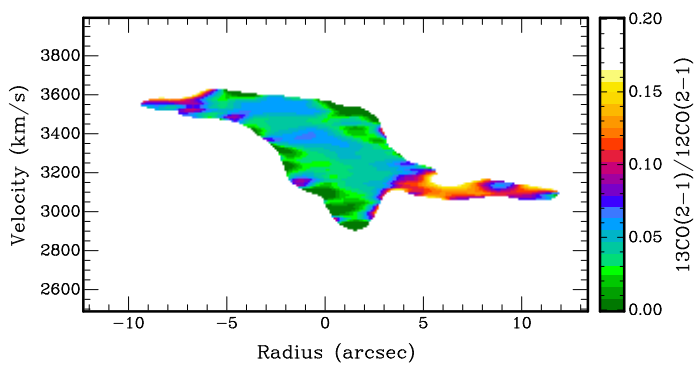}
\caption{Position-velocity diagram of the ratio of the low-resolution \dtco\ and \twco\ brightness temperatures along the jet axis of IC 5063 where the data, before taking the ratio, have been averaged over 1.5 arcsec perpendicular to the jet axis.  Only pixels for which the error in the ratio is smaller than 0.07 are shown.  For the regions in the diagram where the fast outflow is occurring, the values should be regarded as upper limits since no \dtco\ emission is detected there. The spatial resolution along the jet axis is 1\farcs52.  }
\label{fig:rat13CO}
\end{figure}

Figure \ref{fig:slice13CO} shows the position-velocity diagram of the \dtco\ emission. Here, only emission is detected from the bright, inner region and from the undisturbed outer disk. Figure \ref{fig:rat13CO} shows the ratio of the \dtco\ and \twco\  brightness temperatures for those locations where the error in this ratio is smaller than 0.07. Because no \dtco\ emission is detected from the outflow, for this gas we can only set an upper limit  to the line ratio. The figure shows that there is a clear relation between this line ratio and the kinematics of the gas and that the interaction of the radio jet with the ISM strongly impacts on the optical thickness of the clouds. High values (on average 0.13;  see Sec.\ \ref{sec:radex}) are observed in the outer disk and much lower values in the outflow (below 0.05).  Fairly low values  (around 0.05) are found in the bright inner structure and in the radio lobes for gas at normal velocities. This may indicate intermediate conditions there, or that two components, one with low  and one with high line ratios, are superposed.  

The ratio \dtco/\twco\ depends  on many factors, such as abundance,  turbulence and optical depth effects. In relatively normal early-type galaxies, observed values for \dtco/\twco\ in general cover the range 0.05--0.3 \citep[e.g.][]{Crocker2012,Alatalo2015} where the lower values are found in galaxies where the gas is disturbed by interactions or other sources of turbulence such as star formation. Ratios in the upper part of this range are found  in more quiescent conditions. \citet{Crocker2012} argue that these  observed trends   are mainly driven by optical depth effects. The differences in \dtco/\twco\ we see within IC 5063 mimic the same trend. The very low line ratios (below 0.015) occurring in  the region of the  fastest outflow  confirm the optical thin conditions for this emission.  For the gas in  the bright, inner region, as well as for the gas near the radio lobes, somewhat higher line ratios are found, but the ratios are still low compared to many other early-type galaxies. This may indicate  that the gas is mildly optically thick there, or that we see a superposition of optically thin and thick gas. The much higher ratios in the outer disk or typical of regular gas disks and clearly show  that optical thick conditions are present in the outer region.

\section{Discussion}
\label{sec:discussion}

\subsection{Excitation temperatures}

The results presented above give clear evidence  that a significant fraction  of the molecular gas is strongly affected by the interaction with the radio jet/cocoon, both in terms of the kinematics and in terms of the physical conditions of the gas. In an elongated structure around the radio continuum source, \rdt\ has values larger than 1 and  at many locations, in particular near the radio lobes, \rde\  is even larger than 2 (Figs.\  \ref{fig:intSlices}, \ref{fig:ratSlice} and \ref{fig:ratImage}). The gas that has these high line ratios is also the gas that participates in the outflow. This is quite consistent with the results derived from comparing the CO(4-3) emission of IC 5063 with that of CO(2-1) by \citet{Dasyra2016}. There we found that, in general, the jet affected gas has \rvt\ $  > 1.0$ and at some locations this ratio is even quite larger than this value. Such values suggest high excitation temperatures as well as that the CO emission is optically thin since such ratios are significantly above the maximum value of 1 for optically thick emission. Given that in our data we also see line ratios (much) larger than 1.0,   our data reaffirm the high excitations and that the emission from the jet affected gas is optically thin. The  low values for \dtco/\twco\ we observe  for the outflowing gas also indicate this.

Below, we do a more detailed analysis of the conditions of the molecular gas using RADEX non-LTE modelling, but first we  derive  estimates of the excitation temperature of the outflowing CO gas by assuming local thermodynamic equilibrium (LTE) and that the gas is optically thin. For these assumptions,  the average value of $\langle$\rdt$\rangle$ = 1.25 (for details, see below) for the  gas with the most extreme outflow velocities  indicates an excitation temperature of \texci\  = 29 K, in good agreement with the value of \texci\ = 32 K found in \citet{Dasyra2016} as being characteristic for much of this gas. 

Another finding of \citet{Dasyra2016} is that in some small regions, \rvt\ has  values higher  than 1.25, implying that locally, \texci\ is well above 30 K. The images of \rdt\ presented in Figs.\  \ref{fig:ratImage} and \ref{fig:ratSlice} do not show such locations, but given that these images were computed from information integrated either spatially or in velocity, such small spots may have been averaged away in the process. To access the presence of regions with high line ratios, we have also computed pixel-by-pixel \rdt\ ratios using the original three-dimensional data cubes, resulting in a three-dimensional dataset containing \rdt\ for each 3-D pixel.  We have not done this for line ratios involving CO(1-0) because the signal-to-noise of the emission of the outflowing gas is too low in the CO(1-0) cube. The highest values for \rdt\ we find in this way are around 1.65.  Such values correspond  to \texci\ = 54 K assuming optically thin emission and LTE.  This is in good agreement with \citet{Dasyra2016} where we find \texci\ = 56 K for the fastest outflowing gas.

\begin{figure}
\centering
\includegraphics[width=\hsize,angle=0]{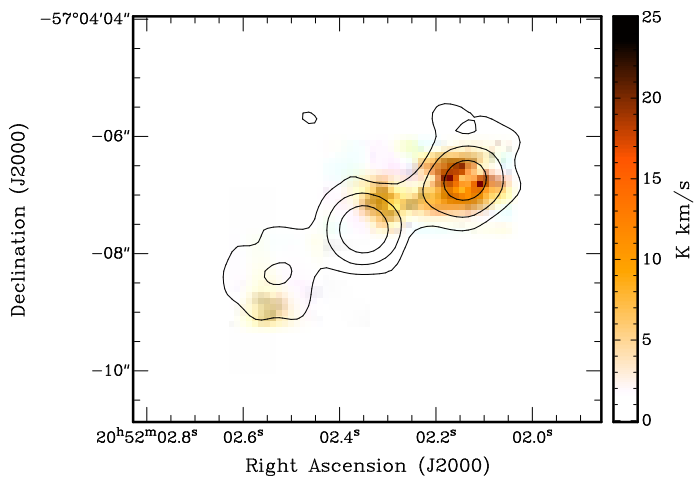}
\caption{ Integrated CO(2-1) intensity using only these pixels in the three-dimensional data cube for which \rdt\ $> 1.55$ and the error in \rdt\ is smaller than 0.5. Contours represents the 346 GHz continuum with  levels 0.15, 0.6 and 2.4 \mJybeam. }
\label{fig:extreme}
\end{figure}

To illustrate the location of the gas with these highest line ratios,  we selected only those pixels in the CO(2-1) data cube for which \rdt\ $> 1.55$   and for which the error in \rdt\ is smaller than 0.5. Using this selection of pixels in the data cube, we produced an image of the integrated CO(2-1) brightness of the emission for which \rdt\ $> 1.55$ and this is shown in Fig.\ \ref{fig:extreme}.
This figure shows that the gas   with highest line ratios is located   near the western lobe and between that lobe and  the core, but, in fact, not coincident with the peak of the W lobe itself. In addition, inspection of the data cubes shows that the outflow velocities of this gas are high, but not as high as the most extreme velocities seen near the W lobe.  The  locations of very high excitation gas we see here with those seen by \citet{Dasyra2016}  roughly coincide.  We note that only a small fraction  ($\sim$5\%) of the total CO(2-1) and CO(3-2) emission comes from gas with this very high excitation. 

\begin{table*}
\label{table:ratios}    
\centering
\begin{tabular}{l r@{$\ \pm\ $}l r@{$\ \pm\ $}l r@{$\ \pm\ $}l r@{$\ \pm\ $}l r@{$\ \pm\ $}l r@{$\ \pm\ $}l   }
\hline\hline
Region & 
\multicolumn{2}{c}{$^{12}$CO(2-1)} & 
\multicolumn{2}{c}{CO(2-1)/} & 
\multicolumn{2}{c}{CO(3-2)/} & 
\multicolumn{2}{c}{CO(3-2)/} & 
\multicolumn{2}{c}{$^{13}$CO(2-1)/} & 
\multicolumn{2}{c}{HCO$^+$/} \\ 
   &
\multicolumn{2}{c}{(K \kms)} & 
   
\multicolumn{2}{c}{CO(1-0)} & 
\multicolumn{2}{c}{CO(1-0)} & 
\multicolumn{2}{c}{CO(2-1)} & 
\multicolumn{2}{c}{$^{12}$CO(2-1)} & 
\multicolumn{2}{c}{CO(3-2)} \\     
\hline
fast outflow &  52.6 & 2.6 & 1.79 & 0.51 & 2.08 & 0.61 & 1.25 & 0.09  & 
\multicolumn{2}{c}{$< 0.015$}  & 0.10   & 0.025 \\
outflow      &  425  & 21  & 1.39 & 0.15 & 1.37 & 0.15 & 1.06 & 0.07 & 0.030 & 0.007 & 0.024    & 0.011  \\
lobes        &  270  & 14  & 1.08 & 0.09 & 0.99 & 0.09 & 0.97 & 0.07 & 0.047 & 0.003 &0.045     & 0.005\\
disk         &  67.7 & 3.4 & 0.40 & 0.03 & 0.17 & 0.01 & 0.39 & 0.03 & 0.13  & 0.01  & \multicolumn{2}{c}{$<0.05$} \\
\hline
\end{tabular}
\caption{Flux integrals and line ratios for the regions defined in Fig.\ \ref{fig:regions} and that were use as input for the RADEX modelling. To incorporate uncertainties in the absolute flux calibration of the different observations, we have assigned a minimum error of 5\% in the flux estimates used \citep{vanKempen2014}.}
\label{tab:ratios}
\end{table*}

\subsection{Mass estimates}

As we remarked in \citet{Dasyra2016}, the fact that the molecular outflow appears to be optically thin affects the estimate of the mass in the molecular outflow compared to the one reported by \citet{Morganti2015} where it was assumed the emission is optically thick\footnote{due to a numerical error, the mass estimate in that paper is about a factor 4 too high}.  To estimate the amount of CO gas participating in the outflow, we have  estimated the flux integral of the outflowing gas by adding the flux from all 3-D pixels in the CO(2-1) cube, and similarly in the CO(3-2) cube, for which \rdt\ $ > 1.0$, taking this limit as a clear indication that the gas is affected by the radio jet.  The flux integrals we find are 9.5  Jy \kms\ for the CO(2-1) line and 25.3 Jy \kms\ for the CO(3-2) line. Given the uncertainties in separating the outflowing gas from the gas with regular kinematics, the flux integral we find for the CO(2-1) lines compares well with  the values of 10.0 Jy \kms\ and 12.3 Jy \kms\ given by \citet{Morganti2015} and \citet{Dasyra2016} respectively. These latter two values  were derived using the kinematics to separate outflowing- and regular gas, instead of line ratios. Both methods likely underestimate the flux of the outflowing gas because, as Fig.\ \ref{fig:ratSlice} shows, at  many pixels in the data cubes, regular and outflowing gas are superposed. Depending on the relative mix of regular and outflowing gas, the measured value for \rde\ can be lower than 1.0 even if a significant component of outflowing gas is present.  To illustrate the effect of this:  lowering  the  value of \rdt\ to separate the two components to 0.9 instead of 1.0, roughly doubles the flux integrals. 

For the optically thin regime, the CO-luminosity-to-$H_2$-mass conversion factor ($\alpha_{\rm CO}$) depends somewhat on the excitation temperature. For $T_{\rm ex} = 29$ K the conversion factor $\alpha_{\rm CO}= 0.25$ K \kms\ pc$^2$ \citep{Bolatto2013}.  Using this conversion factor,  the above flux integrals imply that the  mass of the molecular gas participating in the outflow is $1.3 \cdot  10^6$ \msun\ based on the CO(2-1) flux integral and  $1.1 \cdot  10^6$ \msun\ from the CO(3-2) data. This is about a factor 3 lower than would be derived assuming optically thick emission. Under the assumptions of optically thin emission and LTE,  the mass estimate depends on the excitation temperature \texci. If one uses the highest excitation temperatures we observe ($\sim$55 K), the mass estimate roughly doubles.  However,   the main uncertainty in the mass estimate comes from the uncertainties in the separation of regular and anomalous gas. The results mentioned above from lowering the value of \rdt\ to separate components suggest that the masses reported here are very likely lower limits and the true mass of molecular gas participating in the outflow could be a factor of a few larger.
This means that the cold molecular outflow is about as massive as  the atomic outflow ($3.6\times 10^6$ \msun, \citealt{Morganti2007}), and  much more massive that that of other gas phases \citep{Morganti2007,Tadhunter2014}. Our data clearly show that the impact of the jet on the inner regions of IC 5063 is very significant. However, compared to the mass of the entire ISM of IC 5063 ($\sim$$5\times10^{9}$\msun, \citealt{Morganti1998}), the mass of the outflowing gas is insignificant and the effect of the outflow on the galaxy as a whole is likely to be very limited. 

The mass outflow rate can be estimated using $\dot{M} = 3 v M/R$ \citep{Maiolino2012} where $M$ is the mass of the outflowing gas, $R$ the size of the outflow region and $v$ a representative outflow velocity. This assumes that the outflow has a spherical geometry. Using $R \sim 0.5 $ kpc and $v = 300 $\kms\ gives a   gas mass outflow rate of the about $\sim$12 \msun\ yr$^{-1}$ or somewhat higher. 
This is quite modest compared to many other molecular outflows detected, but similar to the outflow rate  observed in AGN of comparable luminosity as the one in IC 5063 \citep{Fiore2017}.

\subsection{Kinetic temperature, density and pressure}
\label{sec:radex}

Here, and in  \citet{Dasyra2016}, we find that the emission from the fastest outflow  is consistent with arising in optically thin gas. If this gas is in LTE,  it must be warm and dense.  However, in much of the jet-impacted area we find $^{13}$CO(2-1)/$^{12}$CO(2-1)$\sim$0.04. This indicates that the gas is mildly optically thick and therefore we will here perform a non-LTE analysis to investigate the temperature and density for different regions in IC5063. In Fig.\ \ref{fig:sliceHCO} we showed that the CO and HCO$^{+}$ share the same dynamics and thus likely probe the same underlying gas distribution. We will therefore also include the HCO$^{+}$(4-3) line in our analysis.

\begin{figure}
\centering
\includegraphics[width=\hsize,angle=0]{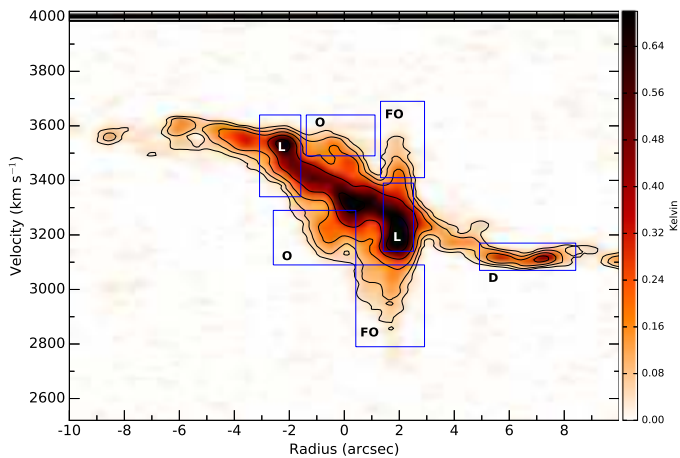}
\caption{Regions for which we computed line ratios as input for the RADEX modelling. The region labelled 'FO' correspond to the fast outflow at the western lobe. Region labelled 'O' to the outflow in the more central regions, while region 'L' to the eastern and western radio lobes at velocities close to those for regular rotation. The region labelled 'D' corresponds to the outer disk.  The underlying gray scale is the position-velocity diagram of the CO(2-1) emission. Contour levels are 43.75, 87.5, 175.0 and 350 mK}
\label{fig:regions}
\end{figure}

\begin{figure*}
\center
 \includegraphics[width=0.49\textwidth]{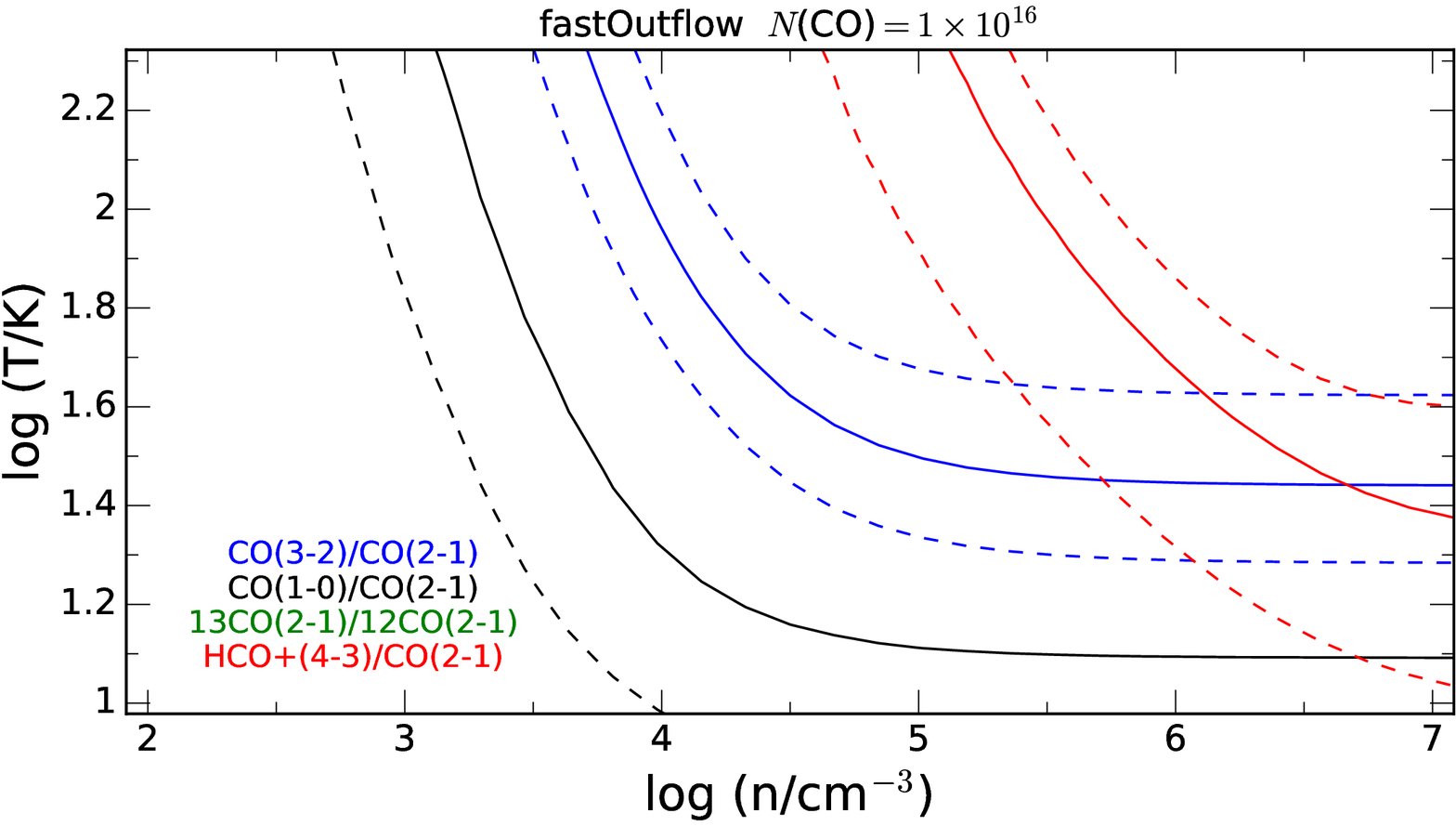}
 \includegraphics[width=0.49\textwidth]{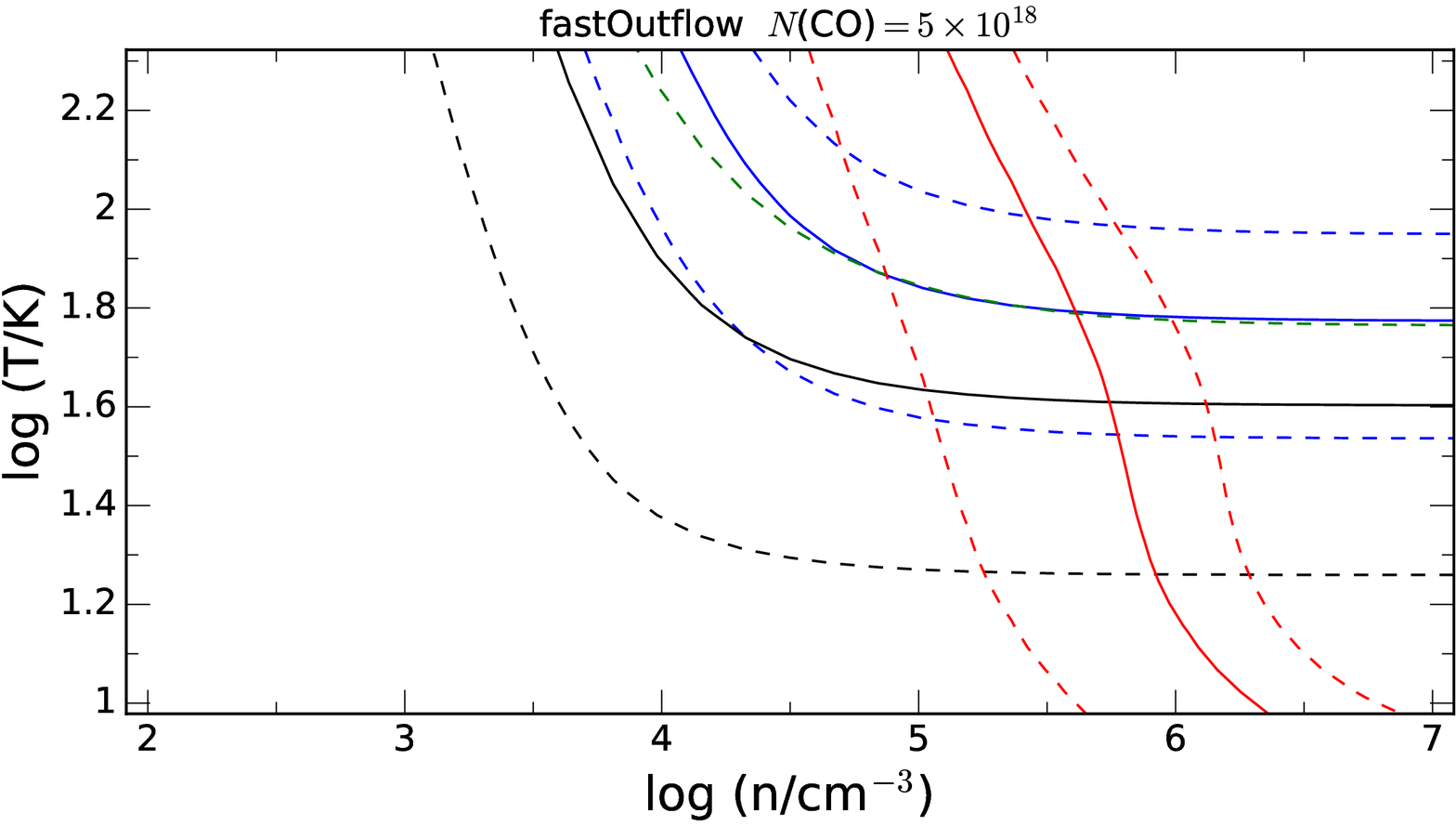} 
 \includegraphics[width=0.49\textwidth]{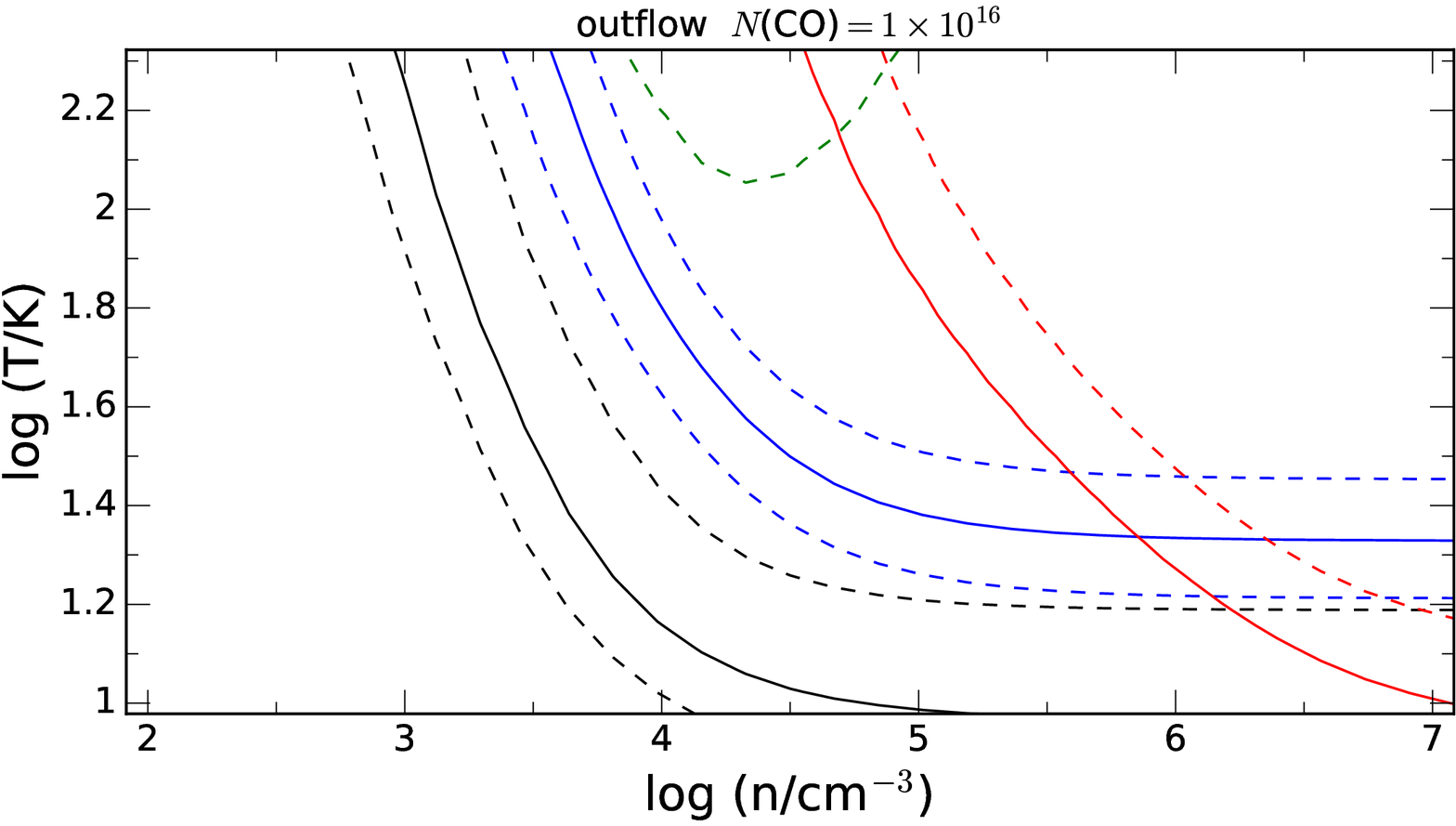}
 \includegraphics[width=0.49\textwidth]{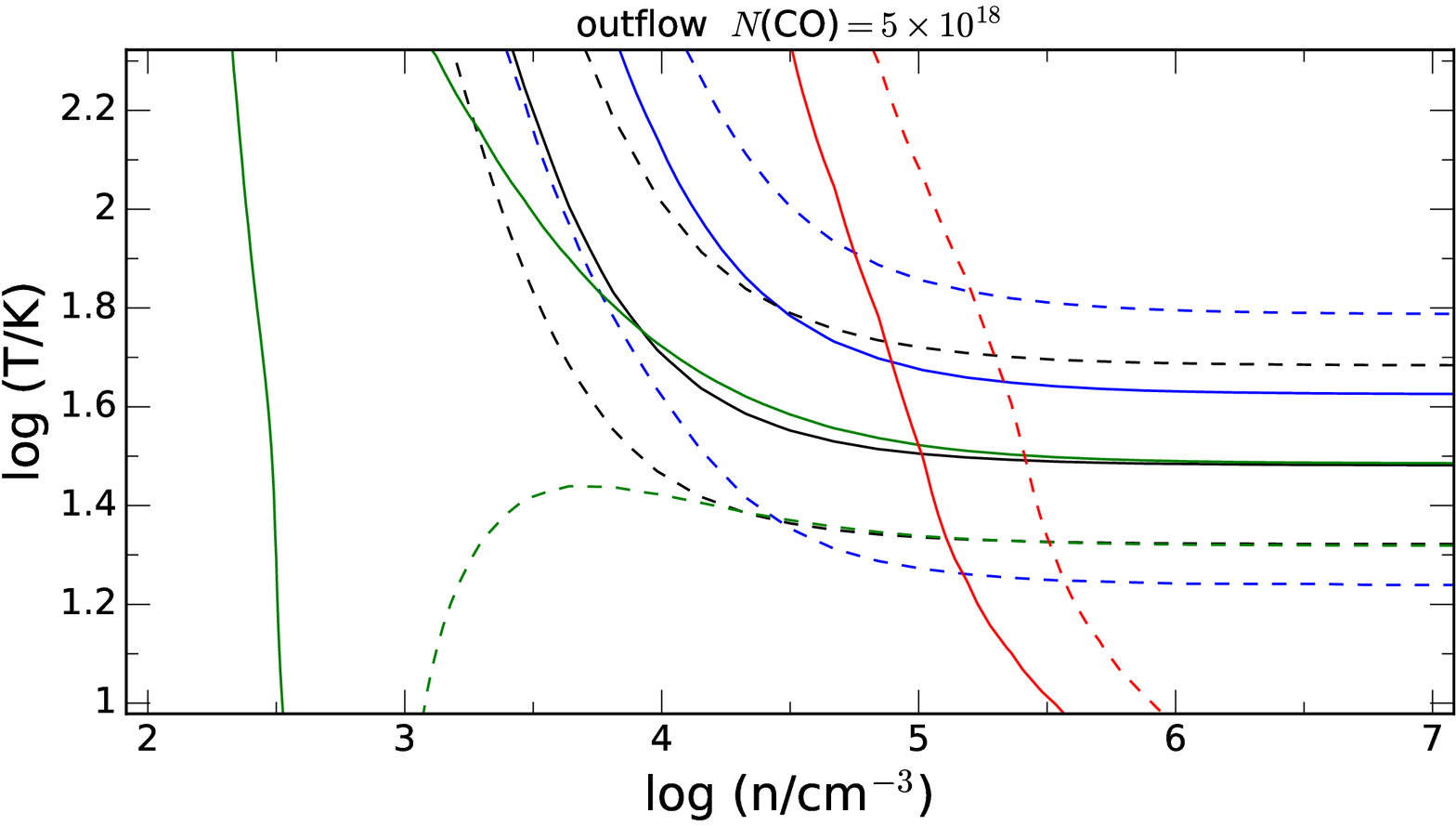} 
 \includegraphics[width=0.49\textwidth]{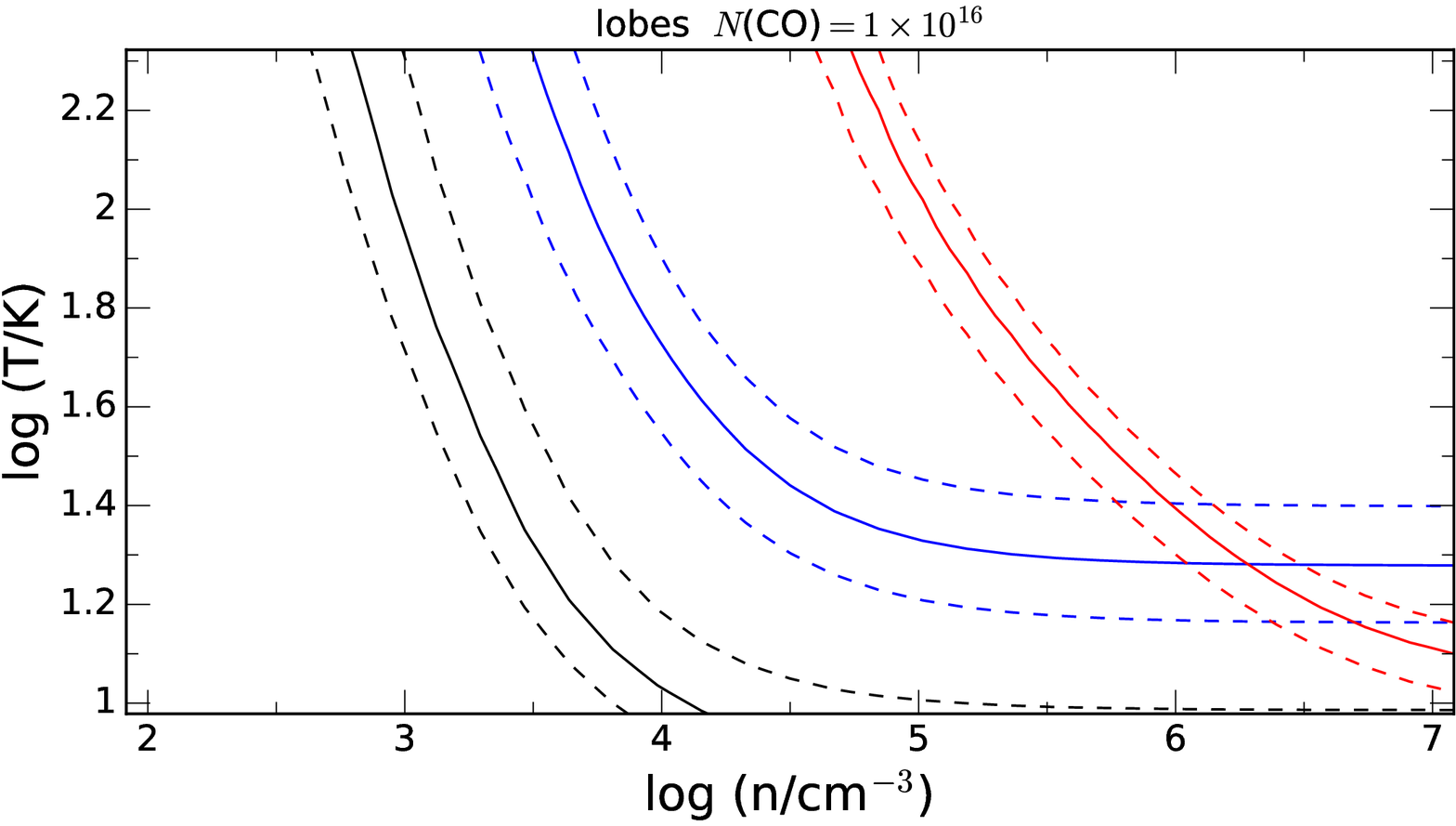}
 \includegraphics[width=0.49\textwidth]{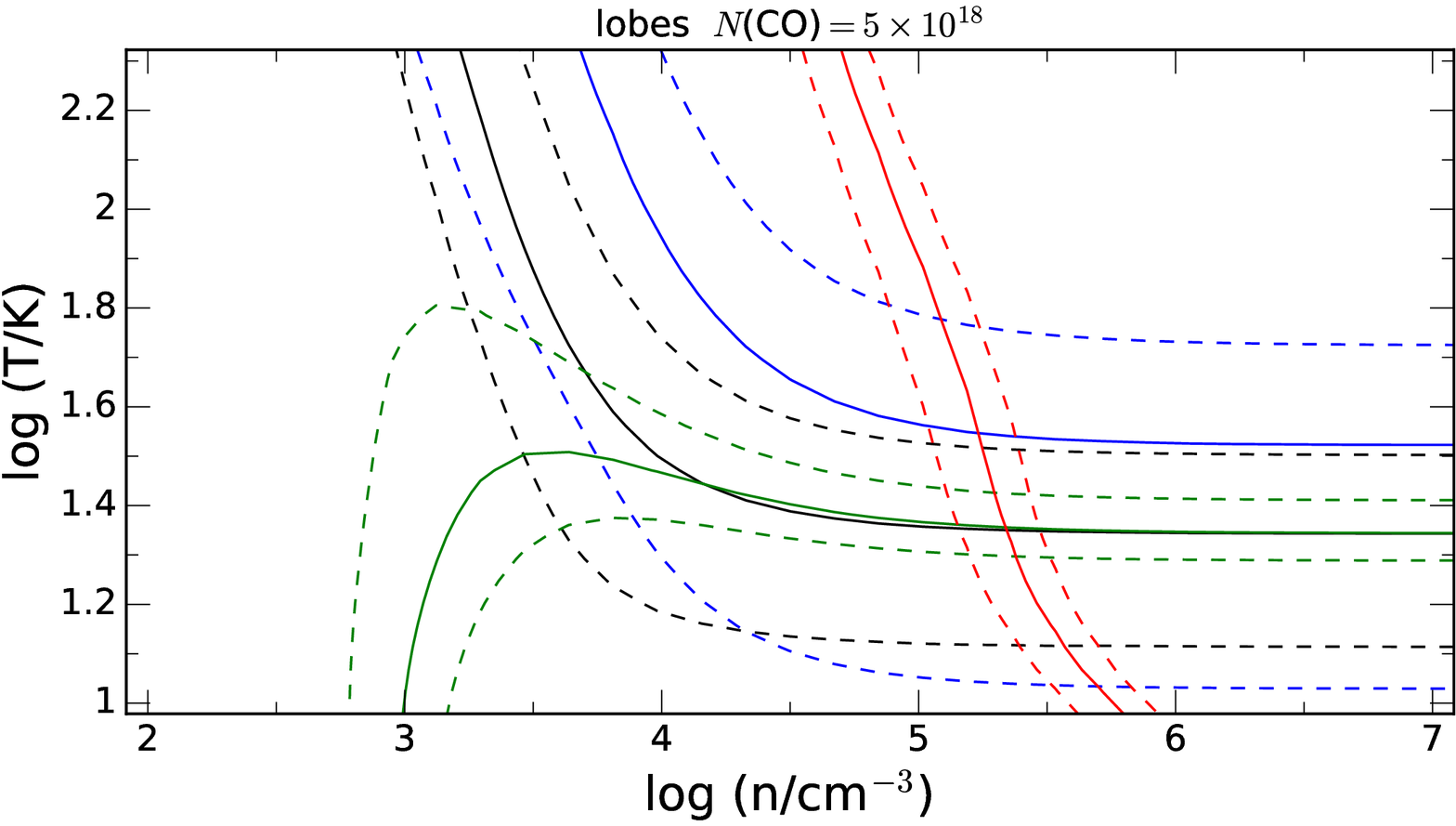}
 \includegraphics[width=0.49\textwidth]{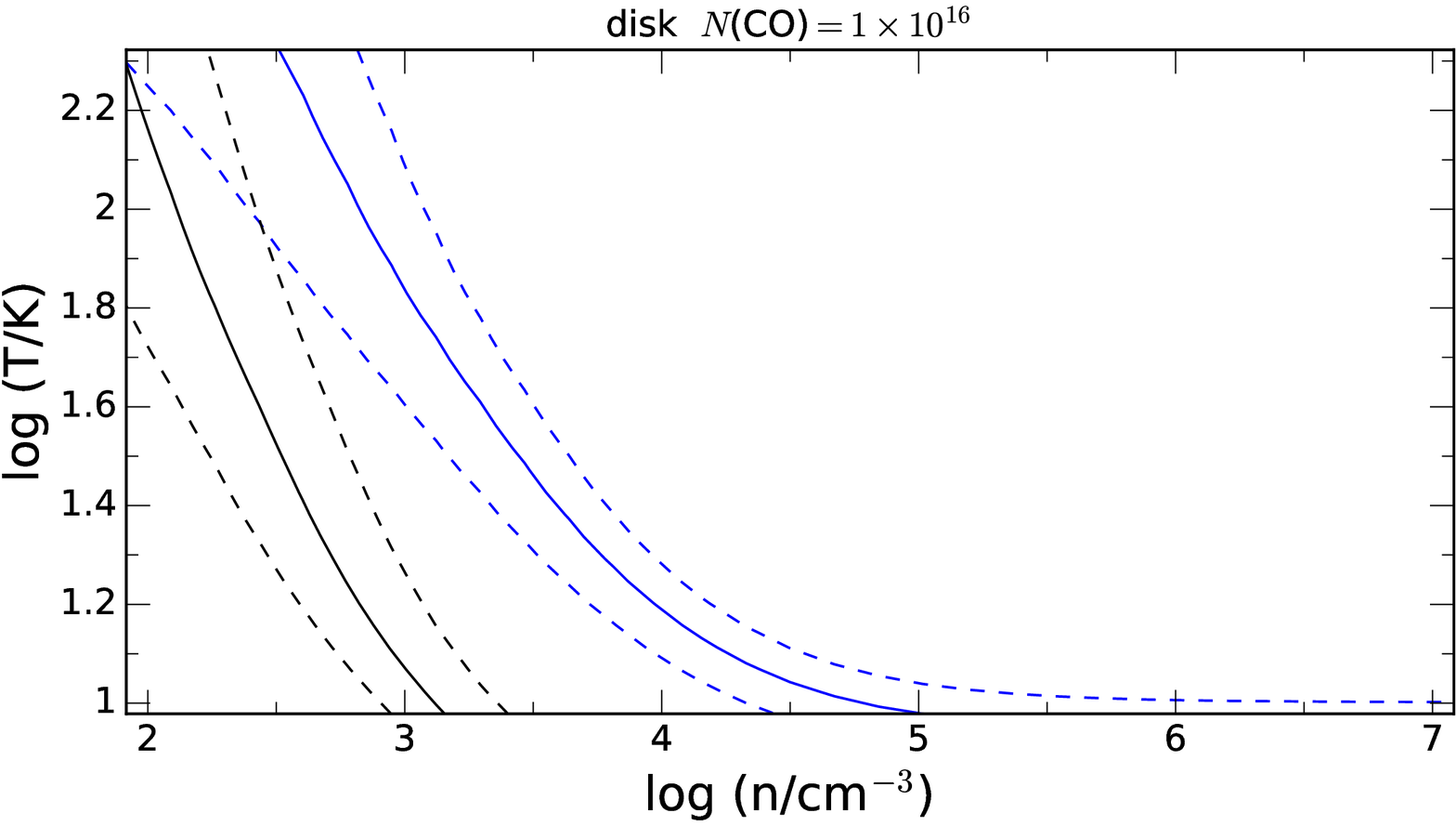}
 \includegraphics[width=0.49\textwidth]{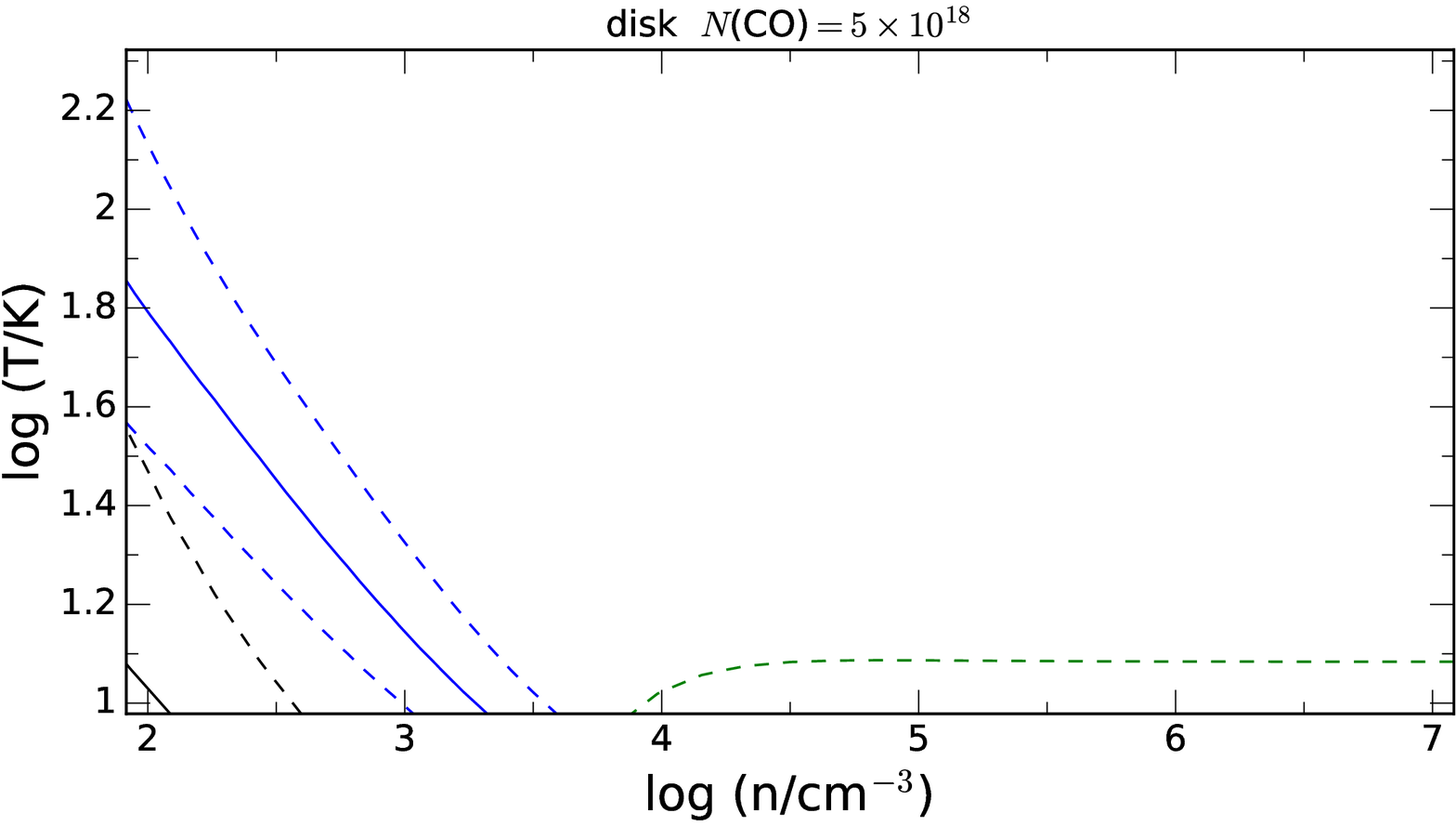}
\caption[]{RADEX models for the four regions (see Fig.\ \ref{fig:regions}). The  constraints on \nhtwee\ and \tkin\  from the various line ratios are indicated (solid lines) and their 3-$\sigma$ errors (dashed lines) for  $N(^{12}$CO) $ = 1 \times 10^{16}$ cm$^{-2}$ (left column) and for  $N(^{12}$CO) $= 5 \times 10^{18}$ cm$^{-2}$ (right column). }\label{fig:f_radex}
\end{figure*}

We use the RADEX non-LTE radiative transfer code to estimate kinetic temperatures and densities for representative regions in the outflow and in the disk of IC5063. Gas heating and excitation models will be presented in a forthcoming paper (Oonk et al.\ in prep.). 
To provide the necessary input for the modelling, we computed line ratios for four representative regions, indicated   in Fig.\ \ref{fig:regions}, covering the various conditions of the molecular gas. The regions defined  include the fast outflow (FO) at the western lobe, the outflowing gas in the more central regions (O), the gas at the locations of the two lobes, but at velocities close to those expected for regular rotation (L), and the gas in the outer disk (D). We defined these regions using the position-velocity diagrams and their layout is shown. For each of the regions, line ratios were computed and results are given in  Table \ref{tab:ratios}. To incorporate uncertainties in the absolute flux calibration of the different observations, we have assigned a minimum error of 5\% in the flux estimates used \citep{vanKempen2014}. 

The very dissimilar  line ratios for the various regions in Table \ref{tab:ratios} clearly underline the different conditions for the gas in the different regions, in particular the large contrast between the outer disk and the regions affected by the radio jet.  A first, main result that immediately emerges from Table \ref{tab:ratios} is that, while the ratios for the outer disk reflect the normal, quiescent conditions in the ISM in this region, the ratios in the jet-affected regions clearly indicate elevated  temperatures and densities. Moreover, between the different regions affected by the jet, conditions are more extreme for gas with more extreme kinematics.

\begin{table*}
 \centering
  \begin{tabular}{ l  c  c c c   } \hline\hline
region       & $\langle N(^{12}\mathrm{CO})\rangle$ & $\log T_{\mathrm{kin}}$ & $\log n_{\mathrm{H_{2}}}$ & $\log P$ \\       
             & (cm$^{-2}$)                          & (K)                     & (cm$^{-3}$)               & (K cm$^{-3}$) \\              
\hline
fast outflow & $1.6 \times 10^{16}$ & 1.5--2.0 & 4.5--6.0 & 6.5--7.5 \\
outflow      & $1.3 \times 10^{17}$ & 1.3--2.0 & 4.0--5.9 & 6.0--7.2 \\
lobes        & $8.2 \times 10^{16}$ & 1.1--1.8 & 5.0--5.8 & 6.8--6.9 \\
disk         & $3.7 \times 10^{16}$ & 1.0--2.3 & 2.0--4.0 & 4.3--5.0 \\
\hline
  \end{tabular}
 \caption{ RADEX analysis of the various regions in IC 5063 (see Fig.\ \ref{fig:regions}). Column 2 gives the  observed area averaged  $^{12}$CO column density. These area averaged measured column densities are derived  assuming LTE and optically thin conditions and  $ T_{\rm ex} = 40$ K, $n = 10^5$ cm$^{-3}$ and $\Delta V = 100$ \kms. Columns 3, 4 and 5 provide the allowed range in physical conditions derived from single slab RADEX models with column densities of $5 \times 10^{18}$ cm$^{-2}$ or higher for the fast outflow, outflow and lobes region. For the disk region, the range given is for all models, also of lower column density.  }\label{tab:results}
\end{table*}

To attempt to obtain more information on this, we modelled the data using a single slab (1-D) model with constant kinetic temperature and density and use the collision rates from the \rm{LAMBDA database} \cite{Schoier2005}. For each line ratio, we predict values for a common grid in \tkin\ and \nhtwee\ (see Fig.\ \ref{fig:f_radex}). For \tkin, we sample the range from 10 to 200 K in logarithmic steps of 0.043 while for \nhtwee\ we sample the range  10$^{2}$ to 10$^{7}$ cm$^{-3}$ in logarithmic steps of 0.167. We set the background temperature to 2.73~K and the line width to 100 \kms. 

In general,  the derived line ratios depend on the input $^{12}$CO column density. We computed models with column densities in the range from $5 \times 10^{15}$ to $1 \times 10^{20}$~cm$^{-2}$  which we sampled in logarithmic steps of 0.5.   For the ratio $^{12}$CO/$^{13}$CO we have assumed 100. Given the sizeable errors in the line ratios involving $^{13}$CO, varying this ratio between 50 and 200 gives similar results, within the errors.  In general, good measurements of the abundance  of HCO$^{+}$ relative to $^{12}$CO are scarce in the literature and the HCO$^{+}$ abundance in IC 5063 is not known. The    probably  most relevant data are given by \citet{Viti2014} in their study of the circum-nuclear disk surrounding the AGN in NGC 1068.  In this circum-nuclear disk they find a relatively high HCO$^{+}$/$^{12}$CO ratio of $\sim$$10^{-4}$ . We therefore opted to use a similarly high value of $1.3\times 10 ^{-4}$.  Lowering the abundance has little effect because the \hco/CO(2-1) tracks shift to somewhat lower densities, but are still well within the error bars of the tracks of the nominal abundance. Increasing the abundance shifts the tracks to higher densities. This means that the results based on the nominal abundance give  an indication of a lower bound to the gas density.

The results are illustrated in Fig.\ \ref{fig:f_radex} and are summarised Table \ref{tab:results}.  Figure \ref{fig:f_radex}  shows,   for two different assumed $^{12}$CO column densities, the  combinations of \nhtwee\ and \tkin\ that reproduce the various line ratios for the various regions. The quality of models can be assessed by   to what extent the different lines in the panels of Fig.\ \ref{fig:f_radex} intersect in a common region. 
 
The trends and differences visible in Fig.\ \ref{fig:f_radex} are  representative for the full range of column densities we considered (see below). The plots show that from the high signal-to-noise $^{12}$CO measurements alone,  we can obtain  tight constraints on the combination of \nhtwee\ and \tkin, but, as expected,  there is a degeneracy between them in the models. Given the slope of the $^{12}$CO tracks, the  pressure ($nT$) is, however, somewhat  better constrained. In principle, since the dependence of the HCO$^{+}$/$^{12}$CO ratio on temperature and density is very different than that of the pure CO line ratios, one can use it to break the degeneracy, but the uncertainty in the  \hco\ data limits its use somewhat. For the outer disk  it has no added value given the low densities of the ISM there. 

We find that the diagrams for $^{12}$CO column densities below $10^{18}$ cm$^{-2}$ are all very similar to each other and  those shown for  $N(^{12}\mathrm{CO}) = 10^{16}$ cm$^{-2}$ are  representative for this column density range. Here, the  main CO line ratios do not quite occupy the same region of the diagrams, although for the fast outflow region, the CO(1-0)/CO(2-1) ratio does not provide  much constraints. For the outflow and lobes regions, the mismatch between the different tracks may be due to the fact that the observed line emission from these regions is a mix of gas with very different conditions, as discussed above, and modelling the line ratio with a model with only one component is not entirely appropriate. For the fast outflow region,  gas mixing is not a problem. The two main CO lines, together with the \hco/CO(2-1) ratio, do not give a well defined solution. When focussing on the CO(3-2)/CO(2-1) (involving the strongest lines detected) and the \hco/CO(2-1) ratios, the low column density models  indicate gas with high density ($ 10^{5} < n < 10^7$ cm$^{-3}$) and \tkin\ between 20 and 50 K for the jet affected gas, with the higher densities and temperatures in the fast outflow region, but this may be related to the mixing of line emission for the outflow and lobes regions discussed above. The resulting pressures are in the range $10^6$--$10^{7.5}$ K  cm$^{-3}$.

When increasing the column density above $10^{18}$ cm$^{-2}$, the character of the diagrams changes and this is illustrated in the right-hand column of Fig.\ \ref{fig:f_radex}. For the jet-affected regions, the tracks of the CO lines shift to higher temperatures and density, while the \hco/CO track shifts to lower density.  Instead, for the disk region the CO tracks move to lower density and temperature. This is likely an optical depth effect with the high-excitation gas being optically thin ( according to the RADEX models with    optical depth in the range $10^{-4}$ -- $10^{-3}$) and the disk gas optically thick.  In particular, the tracks of the main CO lines in the diagrams of the jet-affected regions become more consistent with each other and the models somewhat better define a solution and therefore   the models with high CO column densities in the jet affected regions, in the range  $ 5\times10^{18}$ cm$^{-2}$--$ 5\times10^{19}$ cm$^{-2}$, could be preferred. This may indicate that the true $^{12}$CO column densities in the outflowing gas are at least two orders of magnitude larger than the observed ones, assuming LTE, optically thin conditions, and that the emission fills the beam and which  are in the range $10^{16}$--$10^{17}$ cm$^{-2}$ (see Table \ref{tab:results}). Since the observed column densities are an average over the ALMA beam, this would mean that gas is very clumpy and that the  filling factor of these clumps over the beam is low, at the level of 1\% or lower. The peak $^{12}$CO brightness temperatures observed in our data are around 1 K which  also indicate, for optically thick emission and the excitation temperatures observed, that the gas is clumpy.

The densities resulting from these models are slightly lower than those for the low column density models, but the temperatures are higher and could be as high at 100 K. The  pressures are similar as for the low column density models and are in the range  $10^{6.0}$--$10^{7.5}$ K  cm$^{-3}$. Also for these models, the fast outflow region has somewhat higher density and temperature  (and hence pressure) than the outflow and lobes regions defined in Table \ref{tab:results}. 

For the disk region, since only ratios involving CO lines are available, the models mainly provide  constraints to the combination of \tkin\ and $n$ and not so much to these parameters separately. Nevertheless, it is clear that the densities in the disk  are at least two orders of magnitude lower than in the jet affected regions. Given the slope of the tracks of the CO ratios, the pressure in the disk is better constrained and is in the range $10^{4}$--$10^{5}$ K  cm$^{-3}$ for the low column density models and about an order of magnitude lower for  the high column density models.

Although there are  uncertainties in  the modelling discussed above, nevertheless a fairly clear picture  emerges about the physical conditions of the molecular gas in the outflow in IC 5063 that we can summarise as follows:  it is characterised by increased density and temperature, and  the resulting pressure in the outflowing gas is at least two orders of magnitude higher than in the quiescent ISM of IC 5063. Moreover, the increase in all these physical parameters appears to be strongest for the gas participating in the fastest outflow. 
The much higher densities and pressures of the gas of the outflow would suggest that the gas is compressed and accelerated  by fast shocks driven by the expanding radio source.. The alternative would be that the jets have hit unusually dens molecular material, but in that case the high (precursor) densities would have to be maintained across the 1-kpc region encompassed by the radio components.

\section{Conclusions}

The  ALMA observations we present in this paper of a number of tracers of the molecular gas in IC 5063  give a very detailed, spatially resolved  view of the impact of a radio jet expanding in the ISM of a galaxy, both in terms of the kinematics and in terms of the physical conditions of the ISM. 
 As seen in our earlier ALMA CO(2-1) and CO(4-3) observations \citep{Morganti2015,Dasyra2016} of this object, the new  data of the main $^{12}$CO lines show the fast molecular outflow as a bright inner molecular structure, of about 1 kpc in size and closely associated with the radio jet, embedded in a larger, quiescent regularly rotating gas disk.  Particularly striking is the difference in relative brightness  in the different transitions of this inner molecular structure compared to that of the outer disk, with the inner structure being relatively brighter in the higher-$J$ CO transitions (Fig.\ \ref{fig:totImages}), showing that the gas in the inner molecular structure has very different properties than the gas in the outer disk.

Images and $pv$ diagrams of the line ratios of the main CO lines (Figs \ref{fig:ratImage} and \ref{fig:ratSlice}) clearly show that the outflow has high excitation over a region which is exactly coincident with the radio jet. 
%The excitation correlates with outflow velocity in the sense that the larger the outflow velocities, the higher the excitation conditions are (Table \ref{tab:ratios}). 
These results confirm the model presented in \citet{Morganti2015}, inspired by the simulations of \citet{Wagner2012}, where  the  radio jet is working its way  through a clumpy ISM,  inflating a cocoon in the ISM which is driving the outflow.  

At many locations along the entire jet, the line ratios  \rte, \rde\ and \rdt\ are well above 1.0. This confirms, as  seen earlier in \citet{Dasyra2016} based on \rvt, that the outflowing gas is optically thin and that it has high excitation temperatures.  The \dtco\ data confirm that the outflow is optically thin. The observed line ratios indicate CO excitation temperatures around 30 K for much of the molecular outflow, but at some locations excitation temperatures as high as 54 K are observed, consistent with the results found in \citet{Dasyra2016}.

The inner molecular structure is also detected in  the \hco\ line and   this emission is relatively bright in the fastest outflow in particular. This implies that the densities in the outflowing gas are high. This is consistent with what is observed in other molecular outflows, such as those in, e.g.,  NGC 1068 \citep{Garcia2014} and Mrk231 \citep{Aalto2012,Cicone2012,Feruglio2015}

Estimating the mass of the molecular outflow is complicated by the fact that at many locations in the data cube outflowing and quiescent gas are superposed, making a clean separation between the two impossible. We estimate the mass of the molecular outflow to be at least $\sim$$10^6$ \msun\ and, due to the difficulty in separating outflowing and quiescent gas, it is likely  a factor of a few larger than this value. This means that the molecular outflow is similar in mass as the atomic outflow ($3.6 \times 10^6$ \msun, \citealt{Morganti2007}) and that the cold gas phases dominate the fast outflow in IC 5063 \citep{Morganti2007}.  The  mass outflow rate of the cold gas is $\sim$12 \msun\ yr$^{-1}$. This is lower than observed for many other molecular outflows \citep{Fiore2017} and is likely connected to the fact that IC 5063 is a relatively weak radio AGN.
The mass of the fast outflow in IC 5063 is only a small fraction of the mass of its total ISM  (about $5 \times 10^{9}$ \msun, \citealt{Morganti1998}). The jet-ISM interaction clearly has a significant impact on the ISM in the inner regions of the galaxy, but on the galaxy as a whole, the influence of the outflow will be  much less.

We have explored the physical conditions of the molecular gas using RADEX non-LTE modelling of the observed line ratios in a number of regions of IC 5063. Models with  the outflowing gas being quite clumpy give the most consistent results and our preferred solutions  have kinetic temperatures in the range 20--100 K,  densities ranging between  $10^5$ and $10^6$ cm$^{-3}$ and the resulting pressures being $10^6$--$10^{7.5}$ K cm$^{-3}$, about two orders of magnitude higher than in the outer disk (Table \ref{tab:results}). The higher densities and temperatures are found in the regions with the fastest outflow. The high densities and pressures observed for the outflow suggest that the gas is compressed and accelerated by fast shocks driven by the expanding radio structure.

The results presented in this paper are providing a reference case illustrating the impact of a jet driven by a relatively low-luminosity  AGN
expanding in a gas-rich, clumpy medium.  Spatially resolved observations of outflows driven by higher power AGN are now also needed to build a more detailed picture of the complex processes involved and the impact of the AGN, both on the kinematics and on the physical conditions of the outflowing gas in order to provide realistic constraints for theoretical studies. 

%The above results (the outflowing gas is clumpy and has   much higher temperatures and densities than  gas in a normal ISM) are in line with those obtained for  molecular outflows observed in other AGN (e.g.\ NGC 1068, \citealt{Viti2014,Garcia2014}; Mrk 231, \citealt{Aalto2012,Cicone2012,Feruglio2015}; NGC 1266, \citealt{Alatalo2011}; 4C12.50, \citealt{Dasyra2012} and NGC 1433, \citealt{Combes2013}).  The picture which emerges is that the gas is compressed and fragmented by the mechanism that drives the outflow. 
%
%
%

\begin{acknowledgements}
  This paper makes use of the following ALMA data: ADS/JAO.ALMA\#2012.1.00435.S and ADS/JAO.ALMA\#2015.1.00467.S. ALMA is a  partnership of ESO (representing its member states), NSF (USA), and NINS  (Japan), together with NRC (Canada) and NSC, and ASIAA (Taiwan), in  cooperation with the Republic of Chile. The Joint ALMA Observatory is  operated by ESO, AUI/NRAO and NAOJ.  The research leading to these results has received funding from the European Research Council under the European Union's Seventh Framework Programme (FP/2007-2013) / ERC Advanced Grant RADIOLIFE-320745.

\end{acknowledgements}

\end{document}